\documentclass[aps,prx,twocolumn,showkeys,10pt]{revtex4-2}
\pdfoutput=1 
\usepackage{booktabs}
\usepackage{makecell}
\usepackage{amsmath,amstext,amssymb}
\usepackage[T1]{fontenc} 
\usepackage{xcolor}
\usepackage{soul}
\usepackage{comment}
\usepackage{graphicx}
\usepackage[normalem]{ulem}
\usepackage{hyperref}
\usepackage{multirow}

\graphicspath{{figures/}}

\makeatletter
\def\@bls@top@rule{\vspace{10pt}}%
\def\@bls@bot@rule{\vspace{10pt}}%
\makeatother

\begin{document}

\title{Data-Driven Trends and Subpopulations in the Gravitational Wave Binary Black Hole Merger Population with UMAP}

\author{A.~J.~Amsellem}
\email{aamselle@andrew.cmu.edu}
\affiliation{McWilliams Center for Cosmology and Astrophysics, Department of Physics, Carnegie Mellon University, Pittsburgh, PA 15213, USA}

\author{I. Maga\~na~Hernandez}
\affiliation{McWilliams Center for Cosmology and Astrophysics, Department of Physics, Carnegie Mellon University, Pittsburgh, PA 15213, USA}

\author{A.~Palmese}
\affiliation{McWilliams Center for Cosmology and Astrophysics, Department of Physics, Carnegie Mellon University, Pittsburgh, PA 15213, USA}

\author{J.~Gassert}
\altaffiliation{University Observatory, 
    Faculty of Physics, 
    Ludwig-Maximilians-Universität München, 
    Scheinerstr. 1, 81679 Munich, Germany}
\affiliation{McWilliams Center for Cosmology and Astrophysics, Department of Physics, Carnegie Mellon University, Pittsburgh, PA 15213, USA}

\begin{abstract}
The rapidly expanding Gravitational-Wave Transient Catalog (GWTC) necessitates the development of model-independent techniques to uncover trends and subpopulations within the binary black hole (BBH) population. We present the first usage of the Uniform Manifold Approximation and Projection (UMAP) algorithm, a novel dimensionality-reduction technique, for the purpose of analyzing BBH mergers in GWTC-3. We show that UMAP, paired with a clustering algorithm, effectively partitions the population into four well-segregated subgroups principally via their primary and secondary mass components along with an outlier event, GW$190521\_030229$. UMAP clearly identifies objects in the ${\sim}10~M_\odot$ buildup in the BBH mass spectrum as their own group with aligned spins and mass ratios of ${\sim}0.2{-}0.7$ while objects in or above the ${\sim}35~M_\odot$ overdensity are all in the same, largest group and display typically lower effective spins as well as larger mass ratios (${\sim}0.5{-}0.9$) on average. With the aid of hierarchical population inference, we interpret these as subpopulations from different formation pathways, consistent with previous findings. We also find a transitional group of a handful of objects with masses in between the aforementioned buildups and broad support for anti-aligned spins. We examine the low-mass UMAP subgroup, which exhibits anti-correlation between the mass ratio and effective spin, and show that it drives such anti-correlation for the entire GWTC-3 sample. Overall, we demonstrate that UMAP is an interpretable, non-parametric framework that can not only be used for visualization but also for probing the astrophysics of the BBH population.\end{abstract}


\maketitle

\section{Introduction}\label{section:intro}

The observation of gravitational waves (GWs) \citep{ligo_first_det} opened a transformative era in the understanding of compact object binaries, providing direct access to studies of their formation and evolutionary pathways. Since that first detection, the LIGO/Virgo/KAGRA (LVK) Collaboration has reported a rapidly growing number of events and released increasingly complete parameter estimates through its Gravitational-Wave Transient Catalogs (GWTC). The most recent installments, GWTC-3 and GWTC-4, contain 90 and 153 events, respectively \citep{gwtc3_cat, gwtc4_cat}, spanning mergers between binary black holes (BBHs), neutron star–black hole systems (NSBHs), and binary neutron stars (BNSs). With BBHs comprising the overwhelming majority of detections, population-level analyses of these systems have become increasingly powerful. Such studies have revealed structure in the BBH mass distribution -- including gaps and localized excesses \citep[e.g.,][]{baxter_2021, gwtc3_pop, farah_2024_things, ignacio_2025_bump, gwtc4_pop, antonini_mass_gap, tong_mass_gap, ray_rebut_mass_gap} -- as well as correlations among key binary properties, such as effective spin and mass ratio \citep{safarzadeh_2020_spins, callister_qchi, zevin_2022}. Complementary approaches that jointly model multiple binary parameters \citep{heinzel_2024} further refine our understanding of the underlying population. Together, these emerging features and trends offer critical insight into the astrophysical channels that produce BBH mergers \citep{bavera_2020, zevin_2021} and underpin cosmological applications with spectral sirens \citep[e.g.,][]{farr_2019, ezquiaga_2022, palmese_review_2025, gwtc4_cosmo, ignacio_specsiren_2025}.

Most population studies are highly model-dependent in the sense that the analysis assumes that each (or at least some) of the event parameters are well-described by an analytic formulation. However, current BBH catalogs have not been conclusive in determining a preferred analytic population parameter description. The lack of a uniquep, referred model has led many population studies to opt for ``non-parametric'' or ``data-driven'' methods instead of analytic formulations. Such analyses eschew the use of analytic forms altogether. Instead, they make use of flexible, data-adaptive methodologies -- including Gaussian mixtures \citep{tiwari_2021_gm, rinaldi_2022_gm}, normalizing flows \citep{ruhe_2022_nflows, leyde_2024_nflows}, advanced binning techniques \citep{heinzel_2025}, and Gaussian processes \citep{ray_2023_gp, farah_nonparam, ray_2025_gp, ignacio_2025_gp, antonini_mass_gap, sridhar_2025} -- that can capture physically meaningful quantities and/or correlations in the event parameter space without requiring commitment to specific analytic forms. These non-parametric methods have the added benefit of being able to analyze the full multi-dimensional parameter space jointly, avoiding the inducement of artificial correlations that can result from parameter marginalization \citep{sridhar_2025}.

Within the context of non-parametric methods, we introduce the Uniform Manifold Approximation and Projection (UMAP) algorithm as a method for deconstructing GW parameter populations. UMAP is a dimensionality reduction technique that utilizes manifold learning and topological data analysis to preserve both local and global structure when projecting high-dimensional data into a lower-dimensional space \citep{mcinnes_2018}. Unlike linear methods such as Principal Component Analysis, UMAP focuses on preserving the neighborhood structure of the data, ensuring that events that are similar in the high-dimensional parameter space remain close together in the low-dimensional space. This approach makes UMAP particularly effective at revealing subpopulations and nonlinear relationships that may be obscured in high-dimensional parameter spaces.

In the field of GWs, UMAP has been used to differentiate between GW waveform glitch types \citep{glitch_umap} and detect outlier signals in GW strain data \citep{marianer_2021}. In astronomy and cosmology more broadly, UMAP has been applied to various problems, including delineating populations of exoplanets \citep{exoplanets_umap}, fast radio bursts \citep{frbs_umap}, and gamma-ray bursts \citep{dimple_2024, zhu_2025}; analyzing structure within dark matter halos \citep{susmita_2024}; and characterizing the nature of quenched galaxies \citep{quenching_umap}. In these examples, UMAP has been used both as a form of unsupervised classification of new data and as a means of distinguishing different regimes within an unwieldy dataset. We utilize UMAP for the latter purpose, and -- to the best of our knowledge -- this work represents the first application of the algorithm to GW populations, using UMAP to sort BBH event samples into different groups. Although not the focus of this work, the techniques we develop could, in principle, be used to classify new BBH events.

The current work demonstrates how UMAP can serve as an exploratory tool for dissecting the growing GW dataset without any prior astrophysical assumptions regarding GW sources. In Section~\ref{section:data}, we describe the GWTC-3 catalog and a set of simulations that are used as inputs to the UMAP algorithm. In Section~\ref{section:methods}, we discuss our tuning of the UMAP algorithm for the purpose of analyzing posterior samples of the GW parameters. In Section~\ref{section:results}, we identify distinct subpopulations in the low-dimensional parameter space, examine which GW parameters are most influential in creating the structure and clustering that characterize these subpopulations, and use the UMAP clustering to examine the relationship between GW parameters. In Section~\ref{section:population}, we examine the UMAP-identified subpopulations using hierarchical Bayesian population inference. In Section~\ref{section:discussion}, we connect our UMAP results to theoretical expectations about the BBH population. In Section~\ref{section:conclusion}, we summarize our findings and provide suggestions for further studies that leverage UMAP for the analysis of the GW population.

\section{Data}\label{section:data}
\subsection{GWTC-3}\label{sub:data-gwtc3}
GWTC-3 contains 90 compact binary merger events with a probability of astrophysical origin ($P_{\rm{astro}}$) exceeding 0.5 as identified across multiple search pipelines throughout LVK observing runs O1, O2, and O3. For population analyses that require a sample with high detection confidence and minimal contamination from terrestrial noise, a false alarm rate (FAR) threshold is usually set. Following LVK population inference standards, we only select events with a FAR less than one per year in at least one search pipeline. We also choose to only analyze BBH events for this first UMAP study. After imposing these selection criteria, we are left with a sample of 69 BBH mergers for analysis.

For each selected BBH event, we utilize the parameter estimation posterior samples released by the LVK collaboration. These samples are generated using the Bayesian inference pipeline \texttt{Bilby} \citep{bilby_paper} that compares waveform model predictions to detector data, yielding posterior distributions describing the GW waveforms. These include the detector-frame component masses $(m_{1,\rm{det}}, m_{2,\rm{det}})$, luminosity distance ($d_L$), dimensionless spin magnitudes $(a_1, a_2)$, the angles between the black hole spin vectors and the orbital angular momentum vector ($\theta_1, \theta_2$), and other parameters, most notably the effective inspiral spin $\chi_{\rm eff}$. For astrophysical population studies, it is typical to work in the source frame. To convert the detector-frame masses to the source frame, we assume the cosmological model measured in \cite{planck_cosmo} to obtain the redshifts, $z$, from the detector-measured luminosity distances. We then relate the detector-frame mass to the source-frame mass via
\begin{equation}
    m_{i,\rm{src}} = \frac{m_{i,\rm{det}}}{1+z}, \quad i = 1,2
\end{equation}
where $i=1$ ($i=2$) indexes the heavier (lighter) of the merging black holes.

The effective inspiral spin parameter $\chi_{\rm eff}$ is a dimensionless measure that captures the mass-weighted projection of the black hole spin vector onto the orbital angular momentum axis. $\chi_{\rm eff}$ is the spin parameter typically best constrained by GW observations. It is defined as
\begin{equation}
\chi_{\rm eff} = \frac{a_1 \cos\theta_1 \, m_{1,\rm src} + a_2 \cos\theta_2 \, m_{2,\rm src}}{m_{1,\rm src} + m_{2,\rm src}}
\end{equation}
with $\chi_{\rm eff} \in [-1, 1]$.

The posterior samples released by the LVK collaboration are conditioned on specific prior assumptions that reflect choices made during the parameter estimation. The prior used in these computations is provided in \cite{common_priors} and takes the form:
\begin{equation}\label{eq:prior}
\begin{split}
p(m_{1,\rm src}, m_{2,\rm src}, \chi_{\rm eff}, z) &\propto p(m_{1,\rm det}, m_{2,\rm det}, \chi_{\rm eff}) \\
&\times d_L^2(z) (1+z)^2 \frac{\partial d_L}{\partial z}.
\end{split}
\end{equation}
The term $p(m_{1,\rm det}, m_{2,\rm det}, \chi_{\rm eff}) \: d_L^2(z)$ represents the parameter estimation prior, which assumes isotropic spin orientations as well as uniformity in detector‑frame masses and comoving volume. The other factors arise from the Jacobian of the transformations between detector-frame and source-frame masses and between luminosity distance and redshift.

GW detectors are not isotropically sensitive across the sky or over the full range of binary parameters, and astrophysical inference therefore requires careful treatment of detector selection effects \citep{finn_1993, schutz_2011}. A standard approach is to simulate a large number of GW signals drawn from a chosen astrophysical population, inject them into representative detector noise, and process the resulting data through the same search pipelines used for real observations. By recording which injections are successfully recovered, one can estimate the selection function and quantify the distribution of detectable sources in parameter space. The LVK collaboration provides such sensitivity estimates in the form of discrete samples of recovered injections drawn from a known reference population. For our GWTC-3 analysis, we make the typical choice to use the \texttt{end\_of\_O3\_BBH} injection set from \citep{gwtc3_zenodo}.

To correct for these effects, each posterior sample is assigned a detection probability, $P_{\rm det}(\theta)$, representing the probability that a merger with parameters $\theta$ would have been observed given the detector network’s selection function. This is essential for inferring the astrophysical population of BBH mergers. The selection function weighting for each sample is evaluated using a neural-network-based emulator for \(P_{\rm det}\) \citep{callister_pdet}. The emulator accepts arrays of parameter samples (masses, spins, and redshift) and computes the probability that an event with those parameters would be detected by the LIGO Hanford, LIGO Livingston, and Virgo instruments during the O3 observing run, appropriately marginalizing or randomizing over unconstrained extrinsic parameters when needed. We note that a single event in our BBH population, GW$190521\_030229$, has primary masses that fall above the emulator's mass threshold of 100 $M_{\odot}$. We therefore only use samples from this event that fall below the mass threshold. The emulator has been trained using the sensitivity estimates released by the LVK collaboration and benchmarked using population inference.

To construct a set of posterior samples suitable for unbiased population inference, it is necessary to rescale the original samples by weights that undo both the initial analysis prior and the selection function. For each sample, the weight is computed as:
\begin{equation}
w_j = \frac{P_{\rm{det}}(m_{1,j}, m_{2,j}, a_{1,j}, a_{2,j}, z_j)}{p(m_{1,j}, m_{2,j}, \chi_{{\rm{eff}}, j}, z_j)}.
\end{equation}
where the index $j$ indexes each event sample and the masses are reported (here and in the rest of this work) in the source frame. For each event, the samples and weights are used to construct a gaussian Kernel Density Estimation fit from which we use rejection sampling to produce 400 new samples that we use for all of our analyses.

\subsection{Simulated data}\label{sub:data-sims}
To validate the robustness of our results obtained from real GW data, we run UMAP analyses on large simulated BBH catalogs. These simulated catalogs are generated by drawing BBH parameters from a known population distribution and injecting the corresponding signals into Gaussian noise realizations following the sensitivity measured during the first three months of O3 \citep{lvk_noise}. Bayesian parameter estimation is performed using \texttt{BILBY} \citep{bilby_paper} with the aligned-spin waveform approximant \texttt{IMRPhenomD} \citep{husa_2016, khan_2016}. Each simulated population contains 276 events, approximately four times the size of the GWTC-3 BBH catalog. We use the aforementioned sensitivity estimate to characterize selection effects. We use two of the simulations in \cite{ray_sims}. These are simulated to follow the preferred GWTC-3 model \citep{gwtc3_pop}. 
More specifically, this model includes a \textsc{powerlaw+peak} primary mass distribution \citep{talbot_2018}, a mass-ratio-dependent pairing function \citep{fishbach_2020}, a merger rate evolution scaling as power-law in $(1+z)$ \citep{fishbach_2018}, and a Gaussian distribution for effective inspiral spins \citep{callister_qchi}. For specifics about the model we refer the reader to Appendix~\ref{appendix:methods}. In the first simulation (Simulation I), no intrinsic correlations are imposed between masses, spins, and redshifts. In the second simulation (Simulation II), there exists an anti-correlation between mass ratio, $q$, and effective inspiral spin, $\chi_{\rm eff}$, that is imposed by making the mean of the Gaussian $\chi_{\rm eff}$ distribution a linear function of $q$ with a negative slope \citep{callister_qchi}. The values of the model parameters used in these simulations can be found in Ref.~\cite{ray_2025_gp} (Section 4; Table 1). All simulation data products are available via the Zenodo release \citep{ray_sims_zenodo}.

\section{Methods}\label{section:methods}
One can imagine at least two approaches to running UMAP on the event data in GWTC-3: 1) Pass some ``privileged statistic'' -- for example, a mean, median, etc. -- from each GW parameter for each event to UMAP. 2) For each event parameter, sample the GW parameter distribution and pass \textit{all} of these samples to UMAP. We ultimately choose to follow the second, sample-based option in order to give UMAP the opportunity to distinguish events based on their posterior distributions more broadly while also capturing parameter uncertainties. Additionally, the UMAP results from the sample-based method tend to be more stable in the face of UMAP's inherent stochasticity.

The posterior estimation for each event provides a litany of parameters that could be passed to UMAP as the high-dimensionality input space. Yet, we expect most of these parameters to be either phenomenologically uninformative (e.g., event on-sky coordinates), only broadly constrained, or redundant with parameters chosen to be in our fiducial input space (e.g., luminosity distance is degenerate with the fiducial redshift parameter). Inclusion of these parameters would both slow down the UMAP algorithm and obfuscate our results by adding structures to our final two-dimensional space that are not of particular interest. With this in mind, we restrict our fiducial high-dimensionality space to four parameters: $m_1$, $m_2$, $z$, and $\chi_{\rm eff}$. We also examine other combinations of input space parameters that include merger mass ratio, $q$, and chirp mass, $M_{\mathrm{chirp}}$, while also excluding some of the fiducial parameters.

The UMAP dimensionality reduction acted upon the aforementioned input space can be optimized via a number of tunable hyperparameters. Upon running UMAP with different random seeds, we notice that some of the smaller structures (in opposition to the larger groups) would not appear consistently in the output space. We therefore set the \texttt{n\_neighbors} hyperparameter to 120 (well above its default value of 15), allowing UMAP to consider more neighbors when constructing the topological representation thereby emphasizing more global structure. We also set the \texttt{min\_dist} hyperparameter to 0.05 (below the default value of 0.1) to allow tighter packing of the points, erasing the fluctuating granular structures. We do explore a number of other choices for these and other hyperparameter with the intention of revealing new insights in the resulting output space. Since, however, other hyperparameterizations led to an output space with either less, unstable, or similar structure than the output of our fiducial hyperparameterization, we choose to streamline our results by only including the fiducial hyperparameterization case. Further hyperparameter optimization will be described in Gassert et al. (in prep.).

Since UMAP is an inherently stochastic algorithm, it is important to determine whether our results are robust to multiple random runs or can at least be reproduced by consistently running UMAP with the same random seed. Although we strive for the first standard, we are careful to acknowledge instances that only achieve the latter standard. As described below, except for two minor instances, the first standard is achieved, and our main conclusions are therefore insensitive to the random seeds. We also note that there exists a supplementary level of stochasticity that stems from the sampling during the re-weighting procedure discussed above. To apply a level of consistency in the face of the inherent randomness of this analysis, we use Hierarchical Density-Based Spatial Clustering of Applications with Noise (HDBSCAN) -- a clustering algorithm tuned to UMAP outputs -- to categorize the groups present in our two-dimensional output spaces \citep{hdbscan}. Although this group categorization could probably be executed effectively with visual analysis, we opt for the consistency provided by applying uniform group definitions.

\begin{figure*}[htbp]
  \centering
    \includegraphics[width=\linewidth]{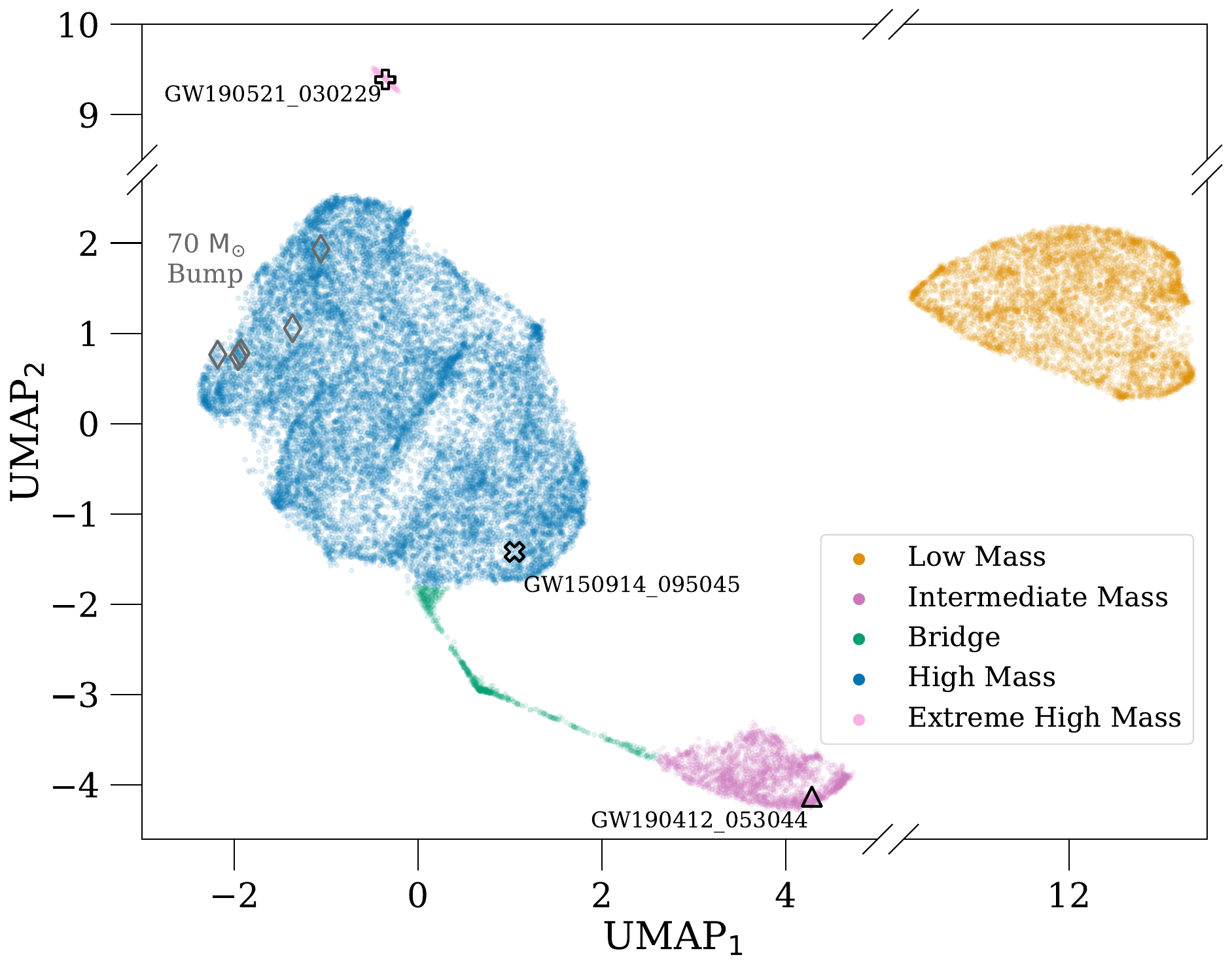}
    \caption{Two-dimensional output space of our fiducial UMAP analysis of GWTC-3 binary black holes. The axes, labeled UMAP$\mathrm{_1}$ and UMAP$\mathrm{_2}$, carry no intrinsic meaning such that relative spacing between points is arbitrary. The fives groups are delineated via a combination of the HDBSCAN clustering algorithm and visual group boundary determination. The median $m_1$ posterior sample for events GW$190521\_030229$ (`+'), GW$190412\_053044$ (`X'), and GW$150914\_095045$ (`$\bigtriangleup$') are also indicated. Lastly, we demarcate the median $m_1$ posterior sample of events that contribute to the ${\sim}70~M_{\odot}$ peak with gray diamonds.}\label{fig:main}
\end{figure*}

\begin{table}
\caption{The 1$\sigma$-level bounds on the primary and secondary mass samples within each group. With the exception of the Bridge group, all groups show strong segregation based on $m_1$ and $m_2$, motivating our mass-based naming convention.}
\label{tab:mass_by_group}
\begin{tabular*}{\columnwidth}{@{}p{0.18\columnwidth}|>{\centering\arraybackslash}p{0.145\columnwidth}>{\centering\arraybackslash}p{0.145\columnwidth}>{\centering\arraybackslash}p{0.145\columnwidth}>{\centering\arraybackslash}p{0.145\columnwidth}>{\centering\arraybackslash}p{0.145\columnwidth}@{}}
\hline
 Parameter & Low & Inter- & Bridge & High & Extreme \\
  & Mass & mediate &   & Mass & High \\
  &   & Mass &   &   & Mass \\
\hline
$m_1 \: [M_{\odot}]$ & $[10, 18]$ & $[20, 31]$ & $[19, 28]$ & $[32, 62]$ & $[88, 94]$ \\
$m_2 \: [M_{\odot}]$ & $[6, 9]$ & $[9, 15]$ & $[12, 19]$ & $[21, 40]$ & $[48, 67]$ \\
\hline
\end{tabular*}
\end{table}

\section{UMAP Results}\label{section:results}

\begin{figure*}[htbp]
  \centering
    \includegraphics[width=\linewidth]{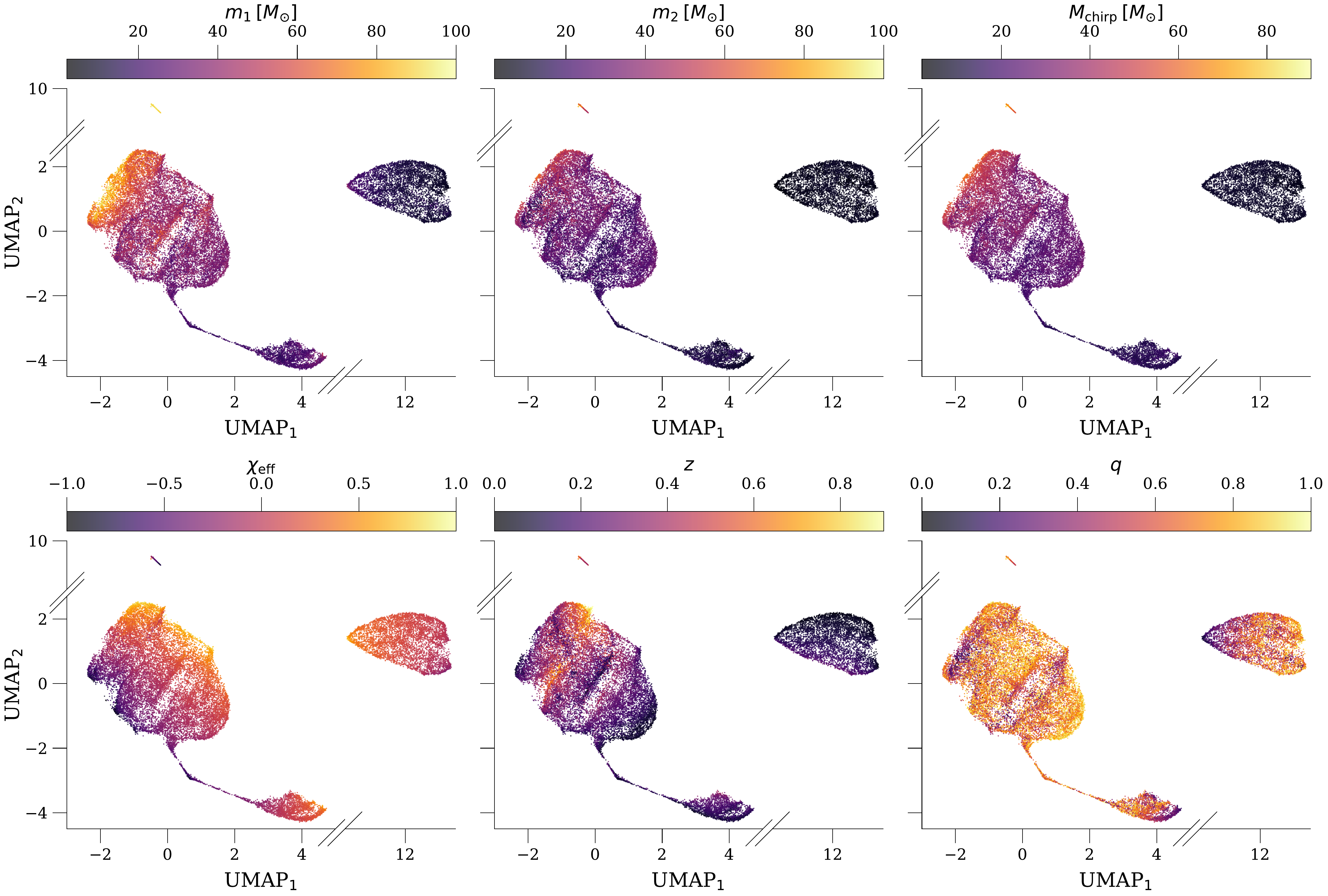}
    \caption{Two-dimensional output space of our fiducial UMAP analysis for GWTC-3 binary black holes, color-coded by the sample's $m_1$ (top left), $m_2$ (top center), $M_\mathrm{chirp}$ (top right), $\chi_{\rm eff}$ (bottom left), $z$ (bottom center), and $q$ (bottom right) value. One can discern increasing and decreasing evolution in these parameters within individual groups.
    }\label{fig:by_param}
\end{figure*}

\subsection{Characterizing the UMAP Groups}\label{sub:umap_groups}
We present the results of our fiducial UMAP run in Figure~\ref{fig:main}, where UMAP$_1$ and UMAP$_2$ represent the two variables in the UMAP low-dimensional space. We consistently observe the BBH event samples cluster into three groups and therefore use the \texttt{HDBSCAN\_flat} function in order to search for exactly three groups. We select the following hyperparameters to be used with HDBSCAN: We use the ``Excess of Mass'' cluster selection algorithm (\texttt{cluster\_selection\_method}); we set the minimum size for each cluster (\texttt{min\_cluster\_size}) to be 15; and we maintain that each cluster point must have a density (\texttt{min\_samples}) of 15 neighboring points. These hyperparameters were chosen such that the three groups are found consistently across stochastic UMAP realizations. Based upon each group's typical $m_1$ values, we refer to the groups as the primary, ``Extreme High Mass'', and ``Low Mass'' groups. We further divide the primary group into the ``High Mass'' and ``Intermediate Mass'' subgroups that are separated by points in the ``Bridge'' group. We perform this latter partitioning by visually determining the extent of the ``High Mass'' and ``Intermediate Mass'' clumps and designating the remaining points to be the Bridge group. We note that the fidelity of the Bridge group is compromised by the fact that, due to the stochasticity of the UMAP algorithm, the group is not always present in the output space. Due to the precariousness of the Bridge group, we are careful to analyze samples near the edges of the group that could have feasibly been included under a different group definition while avoiding making broad claims regarding the group that may not persist in all UMAP realizations or according to different group definitions.

UMAP appears to distinguish groups from each other primarily via the component masses, which motivates our characterization of each group based on its common mass type. Table~\ref{tab:mass_by_group} displays the central 1$\sigma$ ranges of the $m_1$ and $m_2$ distributions within each group, highlighting the discreteness of each group's mass regime. No two groups have any overlap in their 1$\sigma$ mass distributions with the expected exception of the Bridge group. Figure~\ref{fig:by_param} further highlights trends in $m_1$ and $m_2$ within in each group. These masses tend to evolve over some directional axis within each group, yet the exact axis of evolution appears to be different for each group. For example, for the large High Mass group, $m_1$ and $m_2$ seem to decrease as the group is traversed from the top left corner to the bottom right corner. On the other hand, the Low Mass group does not betray an obvious trend in $m_2$ but does display increasing $m_1$ values in the direction of decreasing UMAP$_1$ values. The absence of an $m_2$ trend likely reflects the narrow $m_2$ range in this group (see Table~\ref{tab:mass_by_group}). In the bottom left panel of Figure~\ref{fig:by_param}, one can also discern intra-group trends in $\chi_{\rm eff}$. In the High Mass group, $\chi_{\rm eff}$ clearly increases as the group is traversed from the bottom left to the top right of the group. This axis of evolution appears to be orthogonal to the mass axis of evolution. In the Low Mass group, however, the $\chi_{\rm eff}$ trend appears along the horizontal axis, parallel to the mass trend axis, potentially indicating a stronger correlation between the effective spin and masses in this group compared to the High Mass one. These trends in $m_1$, $m_2$, and $\chi_{\rm eff}$ never display monotonic decreasing or increasing behavior along a given axis, and there are often portions within each group that seem to depart from the stated spatial evolution. Our examination of redshift evolution in the UMAP output space in the bottom center panel of Figure~\ref{fig:by_param} does not reveal  obvious trends. In the Low and Intermediate Mass groups, there is some $z$ evolution perpendicular to the $\chi_{\rm eff}$ evolution, while the trends in the High Mass group are less coherent. Since the redshift range of events in GWTC-3 for a given mass group is relatively small, we are not surprised that we observe less prominent redshift-dependence in the output space. Overall, UMAP appears to mostly reduce our four-dimensional data into two axes relating to component masses and $\chi_{\rm eff}$, but these axes are not consistent across groups and often display deviations from monotonic evolution.

In Figure~\ref{fig:by_param}, we extend the sample coloring to mass ratio, $q$, a parameter that we do not include in our fiducial UMAP input space. Still, one might expect to observe some trends with $q$ in the output space since the parameter is a combination of input parameters $m_1$ and $m_2$. The mass ratio tends to not evolve smoothly over the two reduced UMAP dimensions. When $q$ is passed as an input parameter to UMAP, the mass ratio trends become much smoother (not pictured). Even in the fiducial case, however, we observe that the left portion of the Low Mass group and right portion of the Intermediate Mass group typically harbor low-$q$ samples. Furthermore, in our two population simulations (described further in Section~\ref{sub:sims}), we similarly observe that the low-$q$ portions of lower-mass samples tend to cluster together. As described below, we surmise that these low-$q$ portions of the output space may result at least in part from a parameter-estimation degeneracy between $q$ and $\chi_{\rm eff}$ rather than by a physically different event subsample. We note that the mass ratio trend appears to vary in the same direction as the $\chi_{\rm eff}$ trend for the Intermediate Mass and Low Mass group. We discuss the degeneracy and trend in further detail in Section~\ref{sub:q_chieff}.

Turning to the Extreme High Mass group, we note that it always appears as a satellite group off of the High Mass group in all of our UMAP runs. This group is exclusively populated with samples from GW$190521\_030229$, an event with large $m_1$ and $m_2$, likely indicating a mass-gap progenitor \citep{gw190521}. To determine the degree to which the GW$190521\_030229$ samples can be considered an outlier with respect to the GWTC-3 population, we compute the silhouette score, $s_i$, for the $i$th sample,
\begin{equation}
    s_i = \frac{b_i-a_i}{\rm max(a_i,b_i)}
\end{equation}
where $a_i$ is the mean distance between the $i$th sample point and all GW$190521\_030229$ points and $b_i$ is the mean distance between the $i$th sample point and all non-GW$190521\_030229$ points. A silhouette score close to 1 indicates a sample is closer to GW$190521\_030229$ than non-GW$190521\_030229$ samples while a score of 0 indicates a sample is closer to non-GW$190521\_030229$ than GW$190521\_030229$ samples. Since distances in the output space are highly contingent upon UMAP hyperparameter choices, we note that distance-dependent metrics -- such as silhouette scores -- can be heavily dependent upon choice of hyperparameter. This dependence is mitigated by the fact that relative distances -- such as $b_i-a_i$ -- are not as strongly dependent upon choice of hyperparameters. In our fiducial run, the average of all sample silhouette scores is 0.47 while the average of the GW$190521\_030229$ sample silhouette scores is 0.97. We also observe that only 1\% of samples from GW$190521\_030229$ fall inside the High Mass group instead of the Extreme High Mass group. This silhouette score analysis suggests that the GW$190521\_030229$ samples are separable from all other samples in the output space, justifying our designation of GW$190521\_030229$ as an outlier event. Initially, one might assume that the extreme component masses of GW$190521\_030229$ cause this event to cluster its own distinct group; the analysis in Section~\ref{sub:info_deg}, however, suggests that the $\chi_{\rm eff}$ posterior for this event may also play a role in distinguishing it as an outlier event.

\begin{table*}[htbp]
\centering
\begin{tabular}{l | c | c | c | c | c | c}
\toprule
Event & $m_1 \: [M_{\odot}]$ & $m_2 \: [M_{\odot}]$ & $q$ & $z$ & $\chi_{\mathrm{eff}}$ & $M_{chirp} \: [M_{\odot}]$ \\
\midrule
GW170104\_101158 & 29.8 $\pm$ 4.0 & 20.4 $\pm$ 3.7 & 0.79 $\pm$ 0.20 & 0.18 $\pm$ 0.06 & -0.10 $\pm$ 0.16 & 21.6 $\pm$ 1.7 \\
GW190408\_181802 & 25.2 $\pm$ 2.9 & 17.5 $\pm$ 3.1 & 0.72 $\pm$ 0.15 & 0.19 $\pm$ 0.08 & -0.25 $\pm$ 0.31 & 18.3 $\pm$ 2.0 \\
GW190828\_065509 & 22.3 $\pm$ 4.2 & 12.7 $\pm$ 2.9 & 0.49 $\pm$ 0.21 & 0.17 $\pm$ 0.07 & 0.00 $\pm$ 0.16 & 14.6 $\pm$ 1.1 \\
GW191215\_223052 & 27.6 $\pm$ 3.6 & 19.1 $\pm$ 2.3 & 0.70 $\pm$ 0.15 & 0.23 $\pm$ 0.09 & -0.19 $\pm$ 0.19 & 19.4 $\pm$ 1.3 \\
GW200225\_060421 & 19.7 $\pm$ 2.5 & 13.7 $\pm$ 1.8 & 0.68 $\pm$ 0.16 & 0.13 $\pm$ 0.05 & -0.36 $\pm$ 0.24 & 14.2 $\pm$ 0.9 \\
\bottomrule
\end{tabular}
\caption{Parameter means and 1$\sigma$ uncertainties for events that dominate the Bridge group.}\label{tab:bridge_param}
\end{table*}

\subsection{Specific Events in the UMAP Output Space}
Beyond dissecting the UMAP groups themselves, it is important to examine the characteristics of the specific events that make up each group. For the most part, the samples from each event consistently fall within the same group. However, we have observed that some events have samples in multiple groups (see Appendix \ref{appendix:figures}; Figure~\ref{fig:multigroup}). For all events except five, more than 95\% of the samples fall in a single group while the remaining deviant samples fall in one or more alternative groups. Several additional events, beyond these five, have fewer than 5\% of their samples falling into a group other than the one containing the majority of samples. Which events behave this way varies across different random samplings of the BBH posteriors or across different UMAP runs. The five events that do fall in multiple groups all harbor a significant number of samples in the Bridge group; in UMAP runs in which the Bridge group is not present, samples from each event are less likely to be split across multiple groups. The commonality between these five events is low values of $\chi_{\rm eff}$. We list the parameter values of these events in Table~\ref{tab:bridge_param}. The event that most consistently has non-trivial percentages in multiple groups is GW$200225\_060421$, which harbors the majority of its samples in the Bridge group with nearly all remaining samples falling in the Intermediate Mass group. Adjacent to the samples from GW$200225\_060421$ are samples from GW$190828\_065509$, which also tend to cluster near or in the Bridge group on the side closest to the Intermediate Mass group. The other three events -- GW$170104\_101158$, GW$190408\_181802$, and GW$191215\_223052$ -- tend to cluster near or in the Bridge group on the side closest to the High Mass group. The posterior distributions of these events (and all other events in GWTC-3) are shown in Appendix \ref{appendix:figures}, Figure~\ref{fig:group_breakdown}. Upon examination of the area in and around the Bridge group in Figure~\ref{fig:by_param}, we determine that the placement of these events in the output UMAP space seems to be related to their similarity in primary mass posteriors. GW$200225\_060421$ and GW$190828\_065509$'s lower primary mass posteriors justify UMAP's placement of these samples close to or within the Intermediate Mass group. On the other hand, the higher primary mass distributions of GW$190408\_181802$, GW$190828\_065509$, and GW$191215\_223052$ justify the placement of their samples close to or within the High Mass group. With the exception of these five events, most samples from a given event seem to cluster in one group at a time, suggesting a regular, stable clustering pattern. 

We also examine the output space location of events that are reported to contribute to a peak at ${\sim}70~M_{\odot}$ in the GWTC-3 primary mass spectrum. \cite{ignacio_2025_bump} provide these events (listed in italics in Appendix \ref{appendix:figures}, Figure~\ref{fig:group_breakdown}), and we plot the median $m_1$ posterior sample for each sample in Figure~\ref{fig:main}. Although these events do not visually constitute their own group in the output space, they all tend to cluster together in the upper portion of the High Mass group near the Extreme High Mass group. The only such event that does not fall in the High Mass group is the Extreme High Mass event GW$190521\_030229$. The fact that these events cluster together is not surprising given that the UMAP clustering is driven by primary mass, and these events share similarly large primary masses. Implications regarding the ${\sim}70~M_{\odot}$ peak events will be discussed further in Section~\ref{section:discussion}.

\begin{figure*}[htbp]
  \centering
    \includegraphics[width=\linewidth]{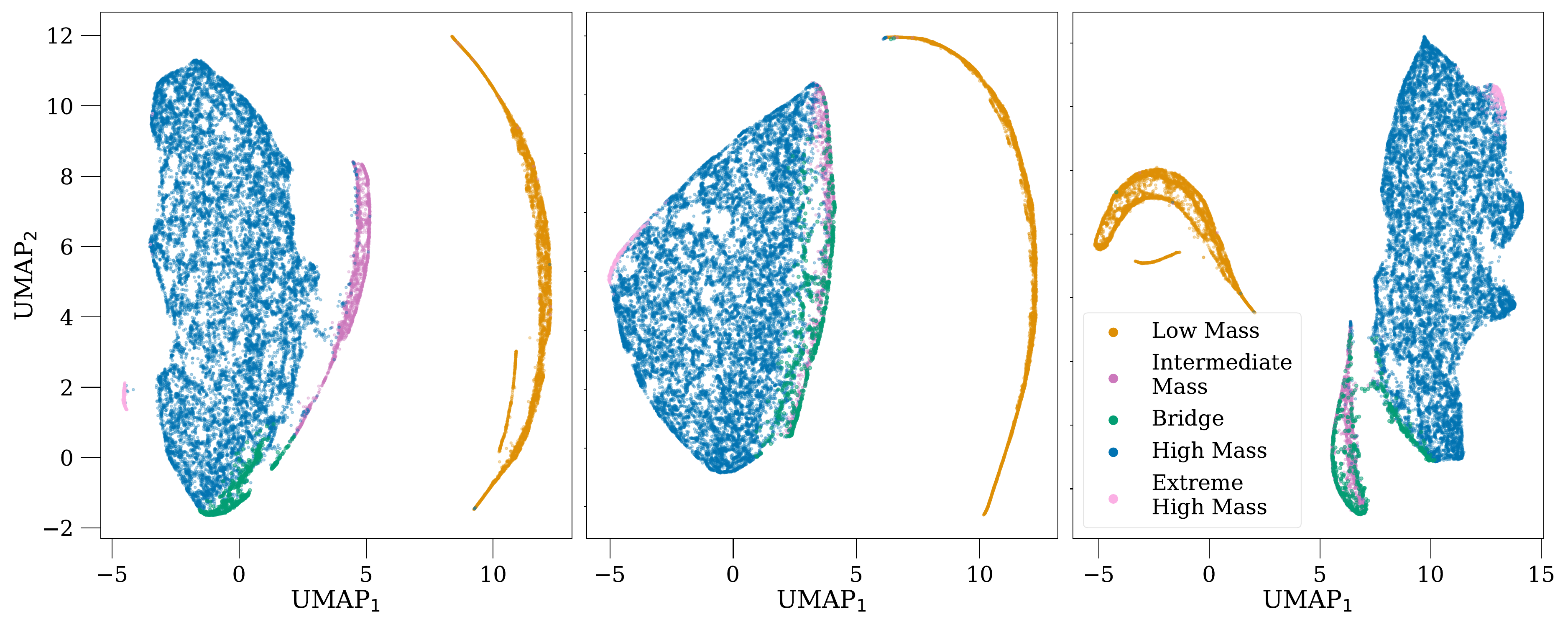}
    \caption{UMAP output space where each panel displays results from a different set of GW parameter inputs into the UMAP algorithm: $M_{\mathrm{chirp}}$ and $\chi_{\rm eff}$ (left), $M_{\mathrm{chirp}}$ and $q$ (center), $M_{\mathrm{chirp}}$ and $z$ (right). For comparison to Figure~\ref{fig:main}, we color each sample by its group designation in our fiducial output space. Most of the fiducial differentiation between samples still appears when only $M_{\mathrm{chirp}}$ and $\chi_{\rm eff}$ are used as inputs.}\label{fig:mchirp_and_params}
\end{figure*}

\subsection{UMAP Input Space}\label{sub:info_deg}
Our fiducial result presented in Section~\ref{sub:umap_groups} utilizes $m_1$, $m_2$, $\chi_{\rm eff}$, and $z$ as the parameters that make up the UMAP input space. We now examine how inclusion and exclusion of specific input parameters change the capacity of UMAP to distinguish between the groups in the fiducial UMAP output space. Since $m_1$ and $m_2$ appear to drive the primary group differentiation in Figure~\ref{fig:main}, we first examine if the information contained in these components can be reduced to a single parameter. We do so by combining $m_1$ and $m_2$ into mass ratio, $q$, and chirp mass, $M_{\mathrm{chirp}} = \frac{(m_1 m_2)^{3/5}}{(m_1 + m_2)^{1/5}}$, and observe whether the original group differentiation is maintained when we pass either $q$ or $M_{\mathrm{chirp}}$ (along with $z$ and/or $\chi_{\rm eff}$) to UMAP instead of $m_1$ and $m_2$. We find that adding $q$ to the fiducial set of input parameters has little effect on clustering in the output space. Moreover, combinations of $q$, $z$, and $\chi_{\mathrm{eff}}$ that exclude $m_1$, $m_2$, and $M_{\mathrm{chirp}}$ fail to produce a group‑differentiated output space. When replacing $m_1$ and $m_2$ with $M_{\mathrm{chirp}}$, however, we mostly recover the original fiducial groups (albeit not in the same spatial configuration) with the exception of the Bridge group. We also notice that including $M_{\mathrm{chirp}}$ while excluding $m_1$ and $m_2$ from the input space leads to the Low Mass group becoming less clumped and much more filament-like. This observation conforms to the following more general trend: the extra information gleaned from incorporating $m_1$ and $m_2$ into the input space adds intricacy to the intra-group structure. Nevertheless, the overall clustering remains primarily governed by chirp mass, which emerges as the dominant parameter driving the UMAP grouping rather than any individual mass or mass asymmetry.

Once we determine $M_{\mathrm{chirp}}$ to be the main parameter of interest, we observe how the addition of $q$, $z$, and $\chi_{\rm eff}$ to the input parameter space affect the UMAP output space. As shown in Figure~\ref{fig:mchirp_and_params}, the inclusion of any one of these parameters allows for some group differentiation. Yet, only when $\chi_{\rm eff}$ is included are all the original groups (again, with the exception of the Bridge group) clearly differentiated. With $q$, one can only distinguish the Low Mass and High Mass groups. The inclusion of $z$ allows the Intermediate Mass group to come into view -- much as the inclusion of $\chi_{\rm eff}$ does --  but it is differentiated from the Bridge group only once $\chi_{\rm eff}$ is added while the Extreme High Mass group is only discerned upon the inclusion of $\chi_{\rm eff}$. The fact that $\chi_{\rm eff}$ is needed to distinguish the Extreme High Mass group -- i.e. the samples from event GW$190521\_030229$ -- demonstrates that the event's differentiation from the High Mass group is not only driven by its extreme masses but by its spin, suggesting that GW$190521\_030229$ is only an outlier once masses \textit{and} spin are considered.  The fact that the Intermediate Mass group can be distinguished from the High Mass group using either $z$ or $\chi_{\rm eff}$ might seem surprising since there does not appear to be an obvious relationship between these two parameters in either the overall GWTC-3 population or in either of these UMAP groups. This result, however, is consistent with the observation that the Intermediate and High Mass groups have distinct distributions in both $z$ and $\chi_{\rm eff}$ (see Figure~\ref{fig:group_hists}). While $M_{\mathrm{chirp}}$ seems to be the primary driver of group differentiation, the addition of either $z$ or $\chi_{\rm eff}$ provides sufficient complementary information for UMAP to fully resolve the Intermediate Mass group as a separate entity.

To summarize, we find it intriguing that $z$, $q$, and $\chi_{\rm eff}$ appear to contain some overlapping information, revealed by the fact that they all are capable of driving at least some amount of group differentiation. We also reiterate that the main GW parameters that drive the UMAP clustering observed in Figure~\ref{fig:main} are chirp mass and effective spin.

\begin{figure*}[htbp]
  \centering
    \includegraphics[width=\linewidth]{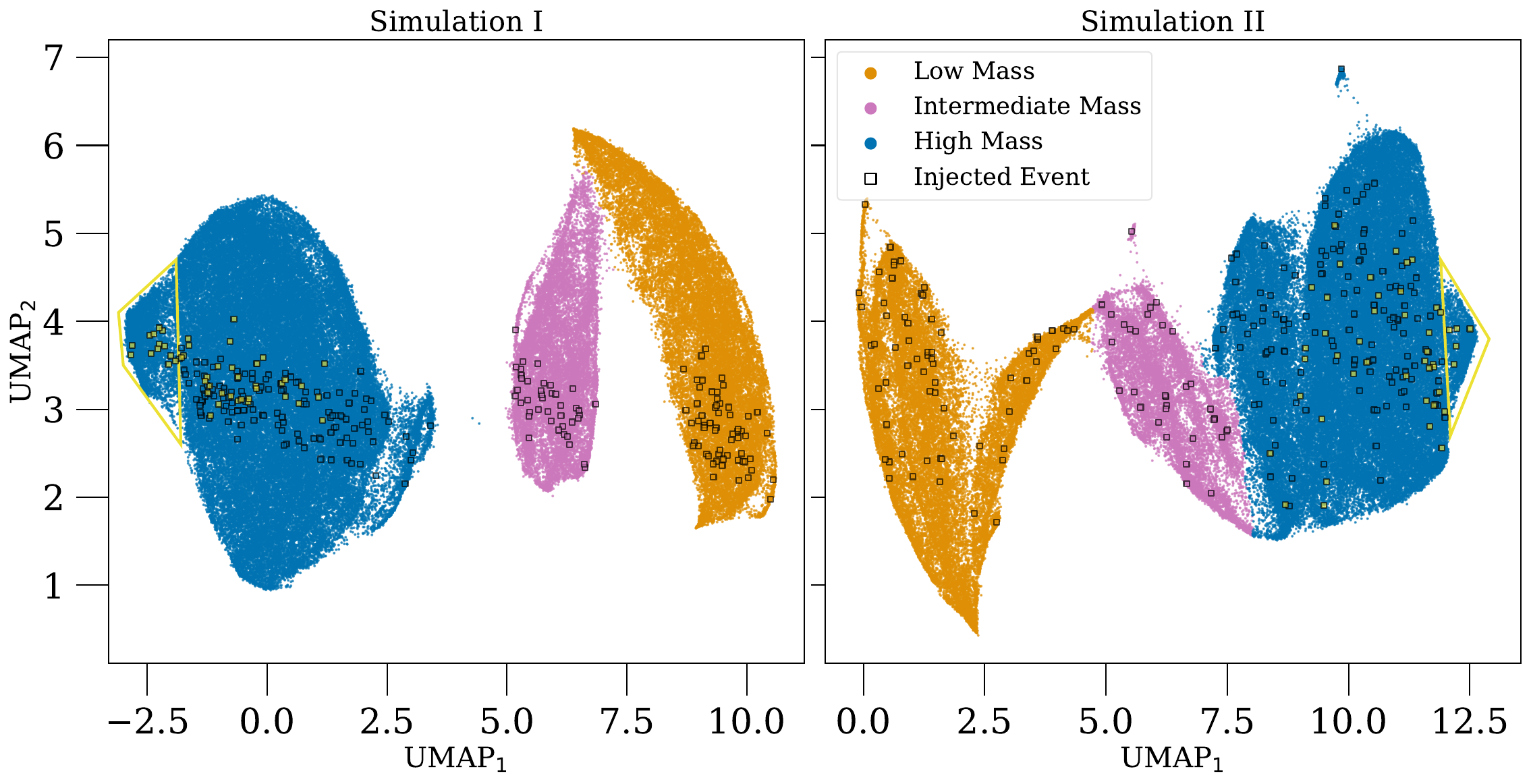}
    \caption{The left (right) panel shows the output space for the Simulation I (II) event samples with each sample colored by its group designation. The black squares represent the true injected parameter values for each event classified in the output space by the embedding produced from the samples. Black squares with yellow fill color indicate injections with $m_1 > 40 M_{\odot}$. The yellow polygons in each panel roughly enclose a region that only contains samples from $m_1 > 40 M_{\odot}$. Both simulations follow a \textsc{PowerLaw+Peak} black hole mass population model, but the main difference between the two is that Simulation II assumes an intrinsic correlation between mass ratio and effective spin. Appendix~\ref{appendix:figures}, Figure \ref{fig:sims_35Msol} displays the clustering of samples near the ${\sim}35~M_\odot$ in the UMAP simulations space.}\label{fig:main_sims}
\end{figure*}

\begin{figure*}[htbp]
  \centering
    \includegraphics[width=\linewidth]{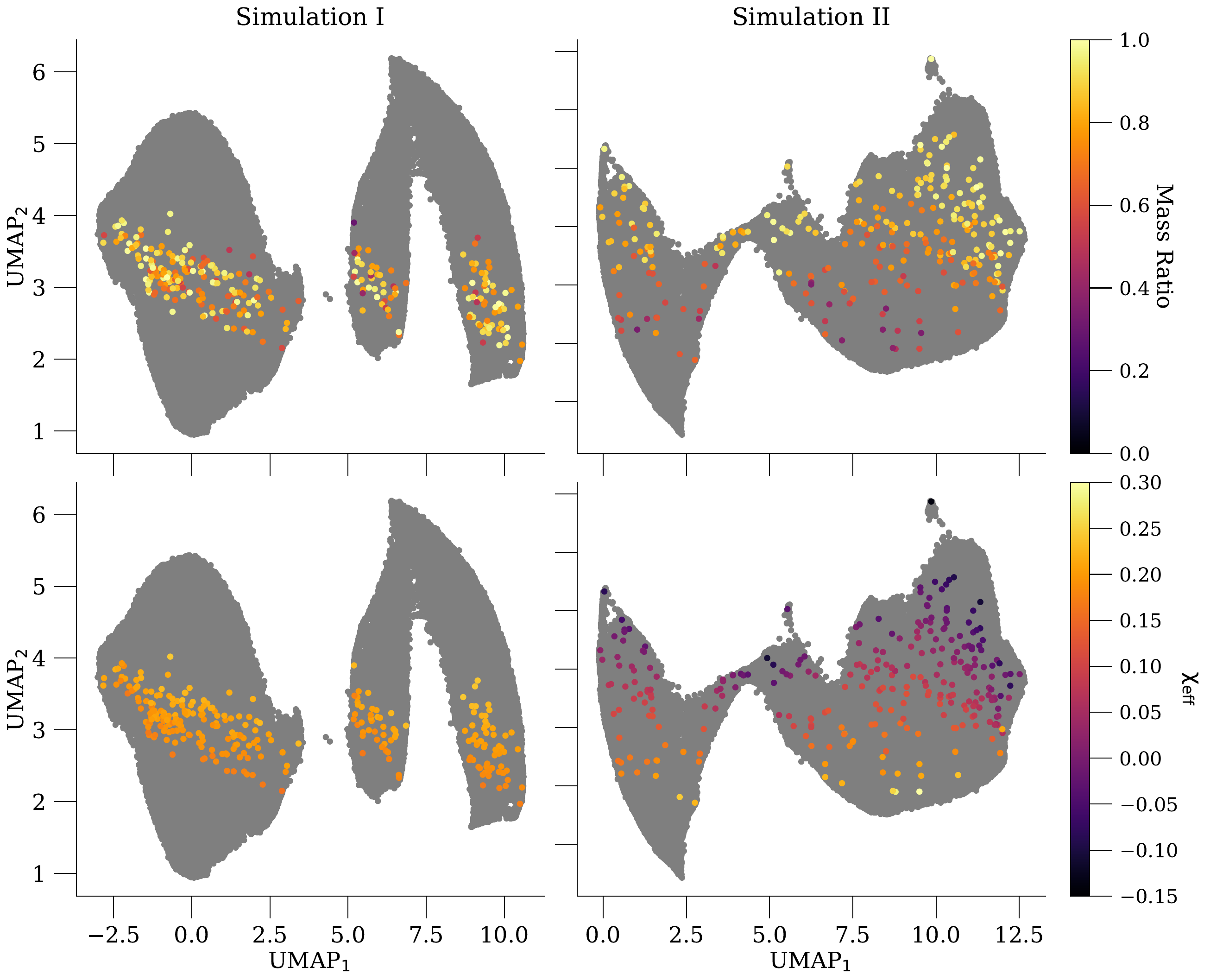}
    \caption{The left (right) panels show the output space for the Simulation I (II) injections with each event colored by its mass ratio (top panels) or effective spin (bottom panels). The gray points represent the UMAP output space of all of the posterior samples of the injected events. Simulation II assumes an intrinsic anti-correlation between mass ratio and effective spin, which is also visually clear in the output space when comparing the top and bottom right panels.}\label{fig:injections}
\end{figure*}

\subsection{Simulations}\label{sub:sims}
We also utilize UMAP on two simulated populations (see Section~\ref{sub:data-sims}) to observe whether they produce an output space similar to the one produced from the data and whether the simulated features and correlations can be identified with UMAP. We examine two sets of simulated events: One in which there exists no $q-\chi_{\rm eff}$ anti-correlation (Simulation I) and one in which it does exist (Simulation II). For each simulated event, we draw 400 random samples from the (much larger set) of event posterior samples. We then weight each sample according to the methodology outlined in Section~\ref{section:data}, redraw 400 samples for each event, and pass these samples to the UMAP algorithm. For our simulation analyses, we keep the UMAP hyperparameters fixed to their default values. 

The general shape of the two-dimensional output spaces appears to be robust to the stochasticity of the UMAP algorithm and the random sampling from the event posteriors. Since the UMAP groups in our simulated populations are less discrete from one another compared to the GWTC-3 data, we adjust the hyperparameters of HDBSCAN in order to consistently identify the same groups over different UMAP realizations. We set the \texttt{cluster\_selection\_method} to ``Leaf'', which allows us to find more clusters. For Simulation I (II), we set the \texttt{min\_cluster\_size} to 400 (4,000) and \texttt{min\_samples} to 20 (500). These choices were found to provide stable groups. Additionally, all samples not assigned to a group are given the same group designation as the nearest sample already in a group. In each population, we request the cluster finder to distinguish three groups (similar to what is found in the data), which we designate as the ``Low Mass'', ``Intermediate Mass'', and ``High Mass'' groups. These groups roughly display parallel $m_1$ values to the original groups of the same names produced in the fiducial UMAP analysis of GWTC-3. 
After running UMAP on the simulated posterior samples, we use UMAP to place each event injection -- consisting of a \emph{true} $m_1$, $m_2$, $z$, and $\chi_{\rm eff}$ -- within the two-dimensional embedding constructed from the samples. The resulting UMAP output space for each population is shown in Figure~\ref{fig:main_sims}.

The output space constructed from the simulations bears some similarity to the output space produced from GWTC-3. We are able to demarcate three groups -- Low Mass, Intermediate Mass, High Mass -- that exhibit mass regimes similar to their counterparts in the GWTC-3 output space. As in the case of GWTC-3, the High Mass group is the largest. In simulations, this group is relatively amorphous with the exception of two  secondary structures emerging at the edges of the group. It is remarkable that these two structures correspond mainly to events above and below the ${\sim}35~M_\odot$ peak (see Appendix \ref{appendix:figures}, Figure \ref{fig:sims_35Msol}), and that the bulk of the group corresponds to events in the ${\sim}35~M_\odot$ peak, showing that UMAP is able to identify a structure that was input in the simulated population. 

By contrast, the low-mass groups tend to exhibit clear curved structure. Additionally, in both simulations, the Low and Intermediate Mass groups are contiguous with one another whereas in GWTC-3 these groups are distinctly separated. We also observe that in each simulation there is a region of the output space -- between the Low and Intermediate Mass groups in Simulation I or \textit{within} the Low Mass group in Simulation II -- where two curved structures meet. This visual difference is reflected in the silhouette scores computed using only samples from the Low and Intermediate Mass groups. In GWTC-3 the mean silhouette scores are 0.93 for the Low Mass samples and 0.89 for the Intermediate Mass samples, indicating strong separation between the two structures in the embedding. The simulations, by contrast, yield much lower values -- Simulation I: 0.51 (Low Mass) and 0.42 (Intermediate Mass); Simulation II: 0.51 (Low Mass) and 0.57 (Intermediate Mass) -- consistent with the impression that the corresponding structures are contiguous and merge into a shared region of the output space. The samples in these regions tend to have small $q$ values. Notably, in both simulated populations, only samples -- not injections -- exist in these regions. Because our injection prescription draws mass ratios from a power‑law distribution that suppresses low-$q$ mergers, the scarcity of low-$q$ mergers is unsurprising. Assuming that the true astrophysical prescription is well-approximated by our injections, we suggest that a parallel phenomenon might be observed in GWTC-3. Indeed, Figure~\ref{fig:by_param} shows that the GWTC-3 output space also has a low-$q$ region on the left side of the Low Mass group and on the right side of the Intermediate Mass group. Although these regions are clearly separated in the GWTC-3 output space, these regions still may be analogous to the conjoined, low-$q$ structures observed in the simulations. We propose that these low-$q$ samples might arise because the mass ratio is intrinsically weakly constrained for low-mass mergers, which produce signals dominated by the inspiral phase \citep{cutler_1994}. In this regime, the gravitational-wave phase evolution is governed at leading post-Newtonian (PN) order by the chirp mass, which enters the waveform phasing at lowest order. The mass ratio contributes only at higher PN orders and is therefore comparatively less well measured \citep{blanchet_2014, mateu-lucena_2022}. As a result, posterior support can extend to smaller $q$ values even when the true mass ratio is closer to unity. This effect on $q$ is related to the known degeneracy between $q$ and $\chi_{\rm eff}$ (as we discuss in more detail in Section~\ref{sub:q_chieff}), which the inspiral signal alone does not fully break. Overall, we demonstrate that UMAP can construct output spaces from GWTC-3 and our simulations that show a number of similarities.

The most noticeable difference between the results from GWTC-3 and our simulations is the lack of a Bridge or Extreme High Mass group. The fact that we do not observe an Extreme High Mass group is not surprising considering neither simulated population contains more than 10 samples with such high component masses ($m_i > 80 M_{\odot}$) and low effective spin ($\chi_{\rm eff} <-0.2$) even though there is reason to expect that such BBHs would exist \citep{fishbach_2022, ignacio_2025_bump}. The lack of outliers in the UMAP results from the simulations (which do not include outliers by construction) provides further evidence in favor of GW$190521\_030229$ truly representing an outlier compared to the rest of the GWTC-3 population rather than being spuriously identified.

The lack of a Bridge group is perhaps more intriguing even though the GWTC-3 sample parameters that typify this group -- negative $\chi_{\rm eff}$ and high $q$ -- have no equivalent in the simulated populations. If the simulated and GWTC-3 populations were alike in most regards with the exception of Bridge-like samples, one could imagine that the Bridge-less simulation samples might lead to a gap in the UMAP space between the High and Intermediate Mass groups. This is the exact behavior we observe from the Simulation I UMAP result. Simulation II, however, shows no gap between the Intermediate and High Mass groups. In fact, Simulation II does not even exhibit any discontinuities between groups in the output space. The fact that UMAP does not introduce a Bridge-like structure in either simulation might suggest that the Bridge group in GWTC-3 represents a real structure rather than a UMAP artifact. 

We also note that the injections in Simulation I are not as widely distributed in UMAP$_2$ as those in Simulation II. We interpret this as a result of the simplicity of the population model used in Simulation I. Most of the relevant structure can be captured by a single parameter, UMAP$_1$, which is directly related to mass and redshift. For the samples in Simulation I, the observed spread in UMAP$_2$ -- the axis that appears to be more closely related to mass ratio and effective spin -- is therefore mostly a result of the uncertainty in the measurement of these parameters. The injections in Simulation I occupy a tighter region of the output space as they are all centered around $\chi_{\mathrm{eff}}{\sim}0.2$ with a standard deviation of $10^{-1.8}$ while a broader range of true values are allowed in Simulation II. 

We also notice that the output space preserves the injected relationship between $q$ and $\chi_{\mathrm{eff}}$ in Simulation II: it is clear from the color map of the right panels in Figure~\ref{fig:injections} in which mass ratio and effective spin are anti-correlated on the UMAP$_2$ axis. As expected, no such trend appears in Simulation I. Although this effect follows directly from the input model, it is not obvious UMAP's dimensionality reduction would preserve this trend. Of course, when looking at real data we do not have access to the ``injections'' and must rely on the posterior samples, which also show this mapping due to the 1.5PN degeneracy described in Section~\ref{sub:q_chieff}. A mapping is also found in the Low and Intermediate Mass groups from the GWTC-3 results (Figure~\ref{fig:by_param}), which is similar to what occurs with the Simulation I UMAP outputs for the same groups. Based on the UMAP output space  alone, we are therefore unable to clarify whether an intrinsic correlation is actually present in the data.

\begin{figure}[htbp]
  \centering
    \includegraphics[width=\linewidth]{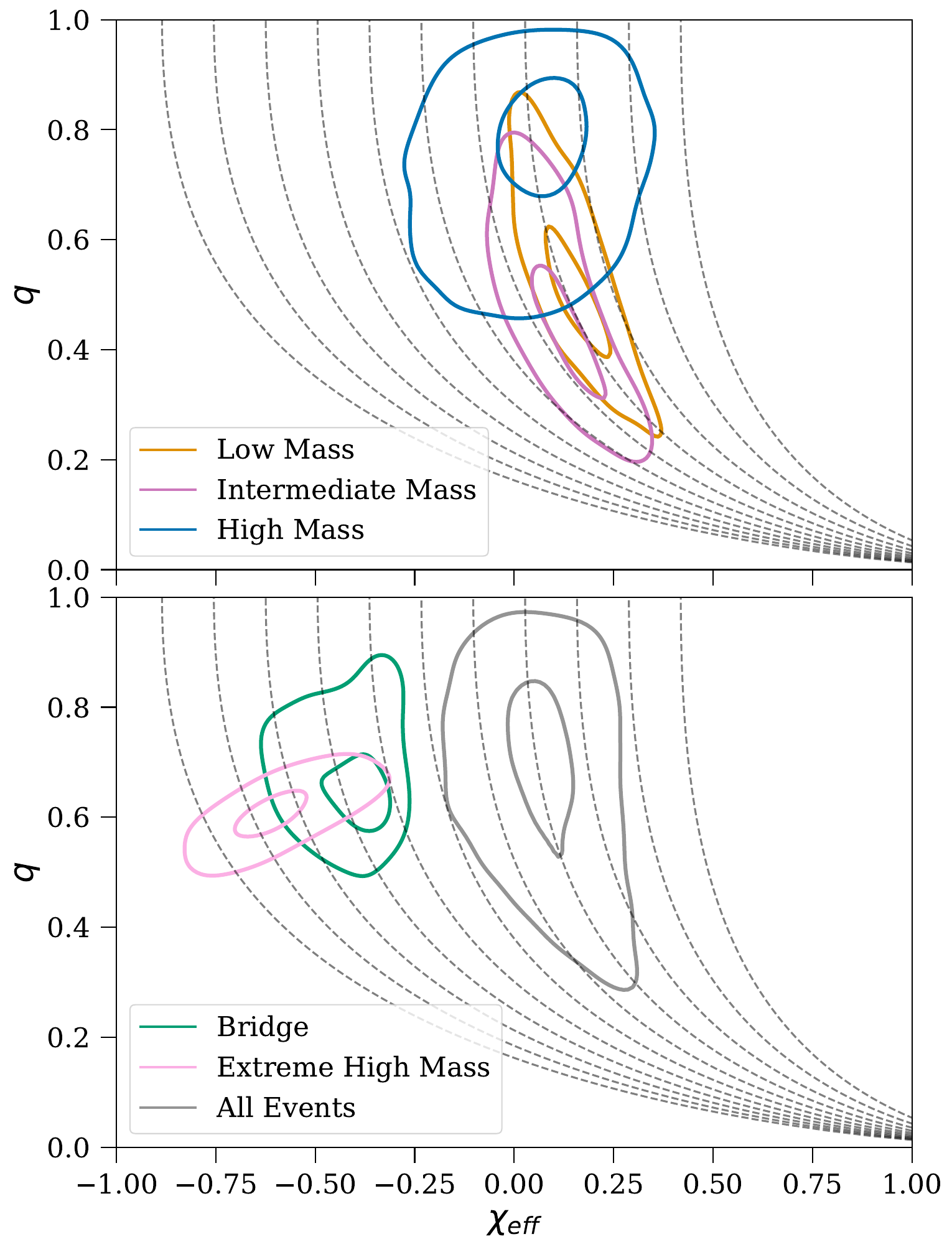}
    \caption{For each group (and all events combined) we construct two-dimensional $q-\chi_{\rm eff}$ Gaussian kernel density estimations of the 50\% and 90\% highest‑density probability contours derived from the samples in each group. Only the low-mass groups show signs of anti-correlation. We also overlay dashed curves indicating the expected constant‑degeneracy tracks from the BBH phase‑evolution parameterization.}\label{fig:q_chieff}
\end{figure}

\begin{table}[htbp]
\centering
\begin{tabular}{l | c | c | c}
\toprule
Group & GWTC-3 & Simulation I & Simulation II\\
\midrule
Low Mass          & -0.66 & -0.80 & -0.75\\
Intermediate Mass & -0.43 & -0.57 & -0.52\\
High Mass         &  0.15 & -0.21 & -0.31\\
All Events        & -0.04 & -0.04 & -0.77\\
\bottomrule
\end{tabular}
\caption{Pearson correlation coefficient computed for samples of $q$ and $\chi_{\mathrm{eff}}$ in different groups and datasets.}\label{tab:pearson_correlation}
\end{table}

\subsection{The \texorpdfstring{$q-\chi_{\mathrm{eff}}$}{q-chi-eff} Relationship in Different Groups}\label{sub:q_chieff}
We now turn our attention to the relationship between mass ratio, $q$, and effective spin, $\chi_{\mathrm{eff}}$, within different UMAP groups. \cite{callister_qchi} note the existence of an anti-correlation between these parameters for 44 BBH candidates in GWTC-2. Pursuant to this observation were a number of theoretical justifications for this trend including modifications to the active galactic nuclei formation channel \citep{mckernan_2022, santini_2023, cook_2024}, the introduction of a subpopulation exclusively consisting of hierarchical mergers \citep{li_2025}, an increase in certain types of stable mass-transfer mergers \citep{olejak_2024}, and/or a preponderance of mergers in which the secondary progenitor becomes more massive than the original primary 
\citep{broekgaarden_2022, banerjee_2024}. \cite{soumendra_2025} suggested that mergers with $m_1<20 M_{\odot}$ are driving the anti-correlation. 

We examine if our groups demonstrate different $q-\chi_{\mathrm{eff}}$ relationships in Figure~\ref{fig:q_chieff}. The posteriors for each group are constructed using a kernel density estimation based on the group samples. Table~\ref{tab:pearson_correlation} displays the Pearson correlation coefficient between $q$ and $\chi_{\mathrm{eff}}$ for each group in GWTC-3 and in each simulation. To test the significance of these coefficients, we perform a permutation-based hypothesis test using random shuffles of the $q$ values while keeping the $\chi_{\mathrm{eff}}$ values fixed. The null hypothesis is that the data are either uncorrelated or positively correlated. We compute the one-sided permutation $p$-value via the equation
\begin{equation}
    p = (N_{\mathrm{ext}} + 1) / (N_{\mathrm{perm}} + 1)
\end{equation}
where $N_{\mathrm{ext}}$ is the number of permutations with a Pearson coefficient less than or equal to the observed value and $N_{\mathrm{perm}}{=}10,000$ is the total number of permutations. To claim that a group exhibits significant anti-correlation, we require the group samples have a distribution with a negative Pearson coefficient and $p{\ll}0.05$. In GWTC-3, the High Mass group displays no signs of anti-correlation between $q$ and $\chi_{\mathrm{eff}}$. The Extreme High Mass and Bridge groups contain too few events, precluding correlation analysis although they appear to occupy a distinct region of parameter space compared to the other groups. In contrast, the Low Mass and Intermediate Mass groups display a visibly clear and statistically significant anti-correlation between $q$ and $\chi_{\mathrm{eff}}$. Their relationship seems to lie close to expected lines of degeneracy that arise from the interdependence of these parameters and the 1.5PN coefficient in the PN expansion for computing the BBH phase evolutions \citep{baird_2013}. One can write the leading-order phase correction in terms of $q$ and $\chi_{\mathrm{eff}}$ as
\begin{equation}\label{eq:1.5PN}
    \Delta \psi_{1.5} = \left(\frac{(1+q)^2}{q}\right)^{\frac{3}{5}}\left[\frac{\left(113 -  \frac{76 q}{(1+q)^2}\right) \chi_{\mathrm{eff}}}{128}- \frac{3 \pi}{8}\right]
\end{equation}
under the assumption that anti-symmetric spin combination, $\chi_{a} = \frac{a_1 \cos\theta_1 - a_2 \cos\theta_2}{2}$, is negligible. \cite{callister_qchi} note that low-mass mergers will display stronger degeneracy since the inspiral parameterizations primarily constrains $q$ and $\chi_{\mathrm{eff}}$ for such mergers. Moreover, we do not expect samples in these groups to follow a single degeneracy curve; rather, \textit{each event} in these low-mass groups has its own $q-\chi_{\mathrm{eff}}$ degeneracy curve determined by the associated merger masses, spins, and signal-to-noise ratio. Although these individual curves are relatively similar -- since they all arise from the same underlying 1.5PN correction -- they are not identical. The observed population-level anti-correlation is therefore, at least partially, a superposition of many slightly different individual degeneracies. Consequently, it is possible that no single degeneracy curve aligns with the population-level anti-correlation pattern. Since we cannot ascertain whether the observed anti-correlation aligns with the expected parameter-estimation degeneracy, and how the selection function may bias the observed population, it becomes difficult to determine if the pattern is a genuine low-mass astrophysical effect or a parameter-estimation artifact. An indication that the relationship may be at least partially artificial is teased out from the fact the groups in Simulation I also show signs of significant negative Pearson coefficients. In fact, all groupings in the simulations -- both with and without an underlying astrophysical $q-\chi_{\mathrm{eff}}$ relationship -- shown in Table~\ref{tab:pearson_correlation} show significant ($p{\ll}0.05$) anti-correlation with the exception of the full Simulation I population, which has a negative Pearson coefficient, but with $p{>}0.05$. One can therefore conclude that the significant anti-correlation displayed by GWTC-3 could result from parameter degeneracies and/or from an intrinsic, physical correlation. To further probe the nature of the $q-\chi_{\mathrm{eff}}$ relationship, we perform a hierarchical Bayesian inference analysis on the subpopulations to further probe the potential anti-correlation in the different subpopulations as well as to model the different spin and mass ratio distributions of the subpopulations.

\begin{figure*}[htbp]
  \centering
    \includegraphics[width=\linewidth]{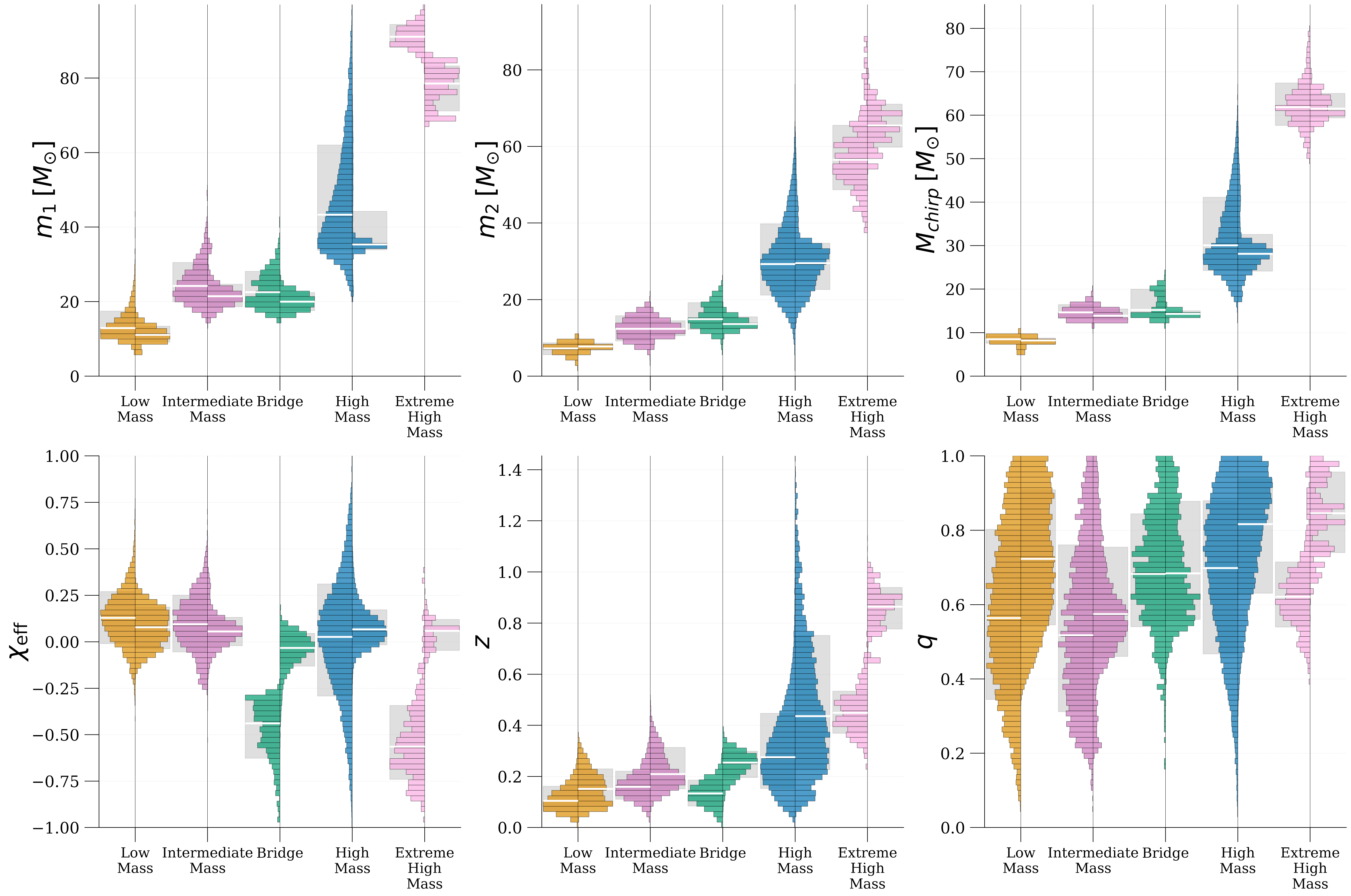}
    \caption{Histograms showing posterior parameter distribution for each group. Each panel displays distributions for a single GW parameter. For each group we display the posterior samples with (right lobe) and without (left lobe) the population reweighting described in Section~\ref{section:population}. Horizontal white lines indicate the distribution medians while gray backgrounds demarcate each distribution's 1$\sigma$ extent. For the non-population-reweighted samples, each sample is assigned to its host group even if most samples from the sample's event fall in another group. For the population‑reweighted cases, samples are grouped according to the group in which the majority of that event’s samples lie.}\label{fig:group_hists}
\end{figure*}

\section{Hierarchical Population Inference}\label{section:population}
We perform a hierarchical Bayesian population analysis of the binary black hole events in the GWTC-3 catalog. Our study considers both the full BBH sample and the sub-populations identified via UMAP. We follow standard choices for the primary mass, mass ratio, redshift, and $\chi_{\mathrm{eff}}$ population distributions as discussed in \ref{sub:data-sims} and used as the fiducial population model for Simulation I and Simulation II. This approach may be seen as counterintuitive given our motivation for a data-driven analysis of the BBH population and given that the standard modeling makes assumptions about the shapes of the parameters distributions. On the other hand, this is the simplest choice to study the \emph{astrophysical} subpopulations, leaving the opportunity for future work to evaluate the subpopulation trends in a fully data-driven manner. Our aim is to focus on the origin of the $q$-$\chi_{\rm eff}$ anti-correlation and identify $q$ and $\chi_{\rm eff}$ distribution differences in the different subgroups. We follow the analysis of ~\cite{callister_qchi} while separately considering the identified UMAP subpopulations. 

\subsection{Statistical Framework}

Because individual gravitational-wave (GW) detections are subject to significant measurement uncertainty and strong selection effects, population-level conclusions are typically inferred through hierarchical Bayesian population frameworks. In this approach, the distribution of true binary parameters $\theta = (m_1, m_2, \chi_{\rm eff}, z)$ is described by a population model $p(\theta \mid \Lambda)$, where $\Lambda$ denotes the set of model parameters characterizing the population distribution. The likelihood of the full catalog is then
\begin{equation}\label{eq:hierarchical_likelihood}
    \mathcal{L}(\{d_i\} \mid \Lambda)
    \propto
    \prod_{i=1}^{N_{\rm det}}
    \frac{
        \int p(d_i \mid \theta)\, p(\theta \mid \Lambda)\, d\theta
    }{
        \int p_{\rm det}(\theta)\, p(\theta \mid \Lambda)\, d\theta
    },
\end{equation}
where $p(d_i |\theta)$ is the single gravitational wave event likelihood for the GW data $d_i$ given measured parameters $\theta$, while $p_{\rm det}(\theta)$ is the probability that a binary with true parameters $\theta$ is detected by the LVK. The denominator in Equation~\ref{eq:hierarchical_likelihood} accounts for selection effects and ensures proper normalization. Posterior samples from individual-event analyses provide Monte Carlo representations of the GW likelihoods, enabling efficient evaluation of the population likelihood. Similarly, the selection sensitivity estimate for GWTC-3 ~\cite{ligo_scientific_collaboration_and_virgo_2023_7890398}, allows for Monte Carlo representations of $p_{\rm det}(\theta)$, enabling the evaluation of the detection efficiency. For a full summary of hierarchical population inference, we refer the reader to Appendix \ref{appendix:methods} and Refs.~\cite{mandel_2019,thrane_2019,vitale_2021}.

\subsection{Population Model}
For our analysis, we adopt a \textsc{PowerLaw+Peak} primary-mass distribution and a merger-rate evolution scaling as power law in $(1+z)$ in accordance with the fiducial choices in the LVK GWTC-3 population analyses \citep{gwtc3_pop}. Following the framework introduced in \cite{callister_qchi}, we employ a two-dimensional spin model that captures linear correlations between the mass ratio, $q$, and the mean and width of the $\chi_{\mathrm{eff}}$ distribution. In particular, the population distribution on $\chi_{\mathrm{eff}}$ is modeled as a Gaussian conditioned on $q$,
\begin{equation}
    p(\chi_{\rm eff} \mid q, \Lambda)
    = \mathcal{N}\!\left(\chi_{\rm eff} \mid \mu_{\chi}(\mu_{\chi,0},\alpha_{\chi}, q), \sigma_{\chi}(\sigma_{\chi,0},\beta_{\chi}, q)\right),
\end{equation}
with its mean, $\mu_{\chi}(\mu_{\chi,0},\alpha_{\chi}, q)$, and its width, $\sigma_{\chi}(\sigma_{\chi,0},\beta_{\chi},q)$, linearly evolving as a function of mass ratio, 
\begin{equation}
    \mu_{\chi}(\mu_{\chi,0},\alpha_{\chi}, q) = \mu_{\chi,0} + \alpha_{\chi} (q - 0.5)
\end{equation}
\begin{equation}
        \log_{10} \sigma_{\chi}(\sigma_{\chi,0},\beta_{\chi},q) = {\log_{10}\sigma_{\chi,0}} + \beta_{\chi} (q - 0.5),
\end{equation}
where $\mu_{\chi,0}$ and $\sigma_{\chi,0}$ are the non-evolving mean and width for the $\chi_{\rm eff}$ population distribution (i.e. when $\alpha_{\chi}=0$ or $\beta_{\chi}=0$, respectively) at a reference mass ratio of $q=0.5$. In summary, the $q$-$\chi_{\mathrm{eff}}$ relationship is parameterized by four population parameters: $\mu_{\chi,0}$, $\alpha_{\chi}$, $\log_{10} \sigma_{\chi,0}$, and $\beta_{\chi}$.

Additionally, compared to the analysis in \cite{callister_qchi}, we use a truncated Gaussian distribution as the population model for the mass ratio $q$,
\begin{equation}
    p(q \mid q_0, \sigma_{q_0}) \propto \mathcal{N}(q \mid q_0,\sigma_{q_0})
\end{equation}
where $q_0$ is the mean and $\sigma_{q_0}$ is the width of the distribution. We ensure that the distribution is valid in the range $[q_{\rm{min}},1]$ and condition on the primary mass to construct the joint $m_1{-}q$ mass distribution. The structure of the $q$ posterior samples in Figure~\ref{fig:group_hists} motivates our use of a Gaussian rather than a more conventional power‑law parameterization. For more details and explicit mathematical relations we refer the reader to Appendix~\ref{appendix:methods}.
\subsection{Results}
\label{sec:results}
We analyze three distinct regions of parameter space, each defined by a chirp-mass interval. Within each region we perform the hierarchical Bayesian inference procedure described above, ensuring that the population likelihood, selection function, and event-level posteriors are all consistently restricted to the corresponding $M_{\mathrm{chirp}}$ domain. The regions of interest are:
\begin{enumerate}
    \item $M_{\mathrm{chirp}} < 10.2\,M_{\odot}$,
    \item $10.2\,M_{\odot} < M_{\mathrm{chirp}} < 17.2\,M_{\odot}$,
    \item $M_{\mathrm{chirp}} > 17.2\,M_{\odot}$.
\end{enumerate}
These intervals correspond to the ranges spanned by the Low Mass group, the Bridge and Intermediate Mass groups, and the High Mass and Extreme High Mass groups identified by the UMAP algorithm. The $M_{\mathrm{chirp}}$ boundaries are also motivated by the fact that the detection probability rises steeply with increasing $M_{\mathrm{chirp}}$. Since each individual event has posterior support confined to a single $M_{\mathrm{chirp}}$ region, the selected boundaries ensure that no events are significantly divided between multiple regions and allow each region to capture a self‑consistent portion of the population in both intrinsic and observed parameter spaces, similarly to what is done in \citep{ray_2025_hiding}. 
For simplicity, we fix all population parameters other than the $\chi_{\mathrm{eff}}$ evolution parameters and the mass ratio distribution parameters to the maximum-likelihood value from the population fit using all of the GWTC-3 events and assuming no mass or redshift evolution for the $\chi_{\mathrm{eff}}$ population distribution. 
We thereby ensure that variations in the inferred $\chi_{\mathrm{eff}}$ distribution are not driven by shifts in the underlying mass and redshift models. The priors on the $\chi_{\mathrm{eff}}$ evolution parameters and the mass ratio distribution parameters are uniformly distributed: $\mu_{\chi,0} \sim \mathcal{U}(-1,1)$, $\log_{10} \sigma_{\chi,0} \sim \mathcal{U}(-1.5,0.5)$, $\alpha_{\chi} \sim \mathcal{U}(-2.5,1)$, $\beta_{\chi} \sim \mathcal{U}(-2,1.5)$, $q_{0} \sim \mathcal{U}(0,1)$ and  $\sigma_{q_0} \sim \mathcal{U}(0.05,0.5)$.

\begin{figure}[t]
    \centering
    \includegraphics[width=0.45\textwidth]{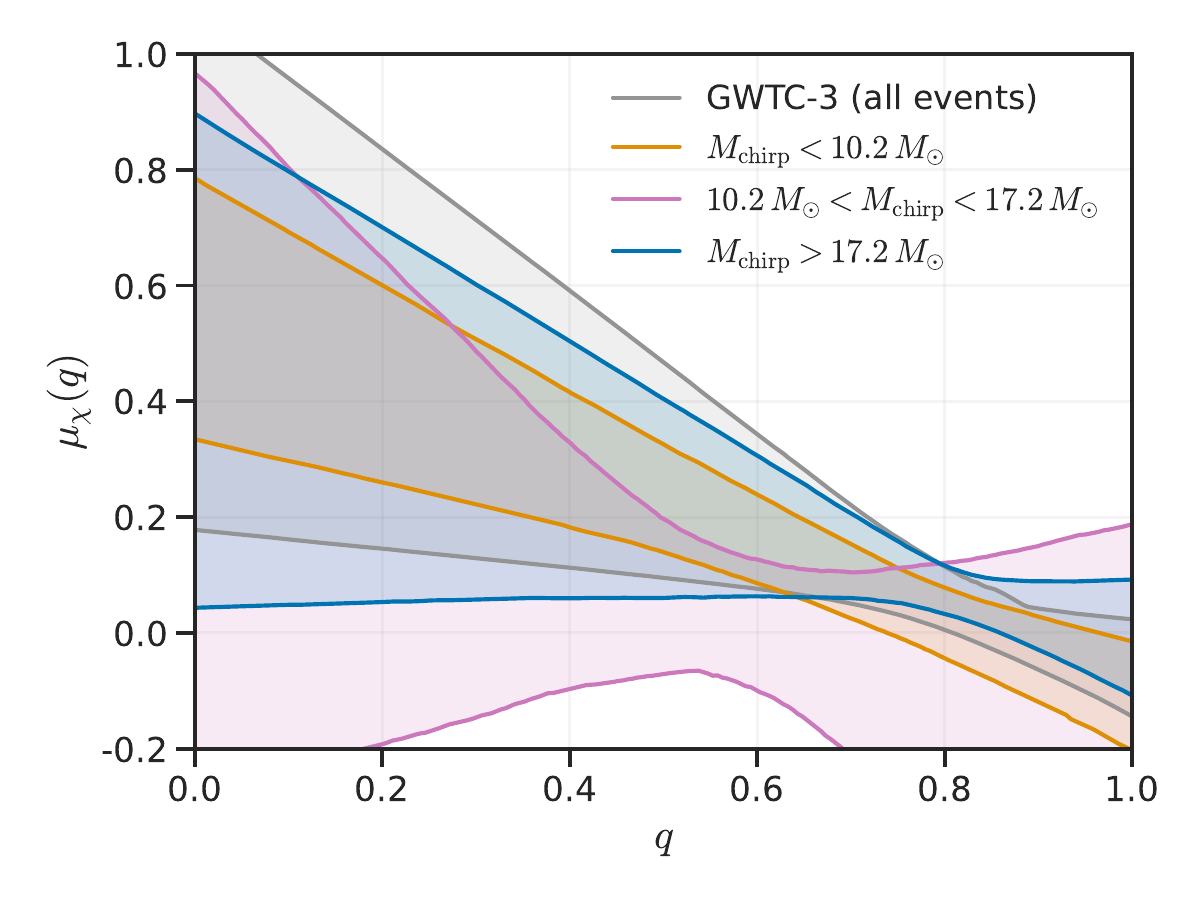}
    \vspace{0.5em}
    \includegraphics[width=0.45\textwidth]{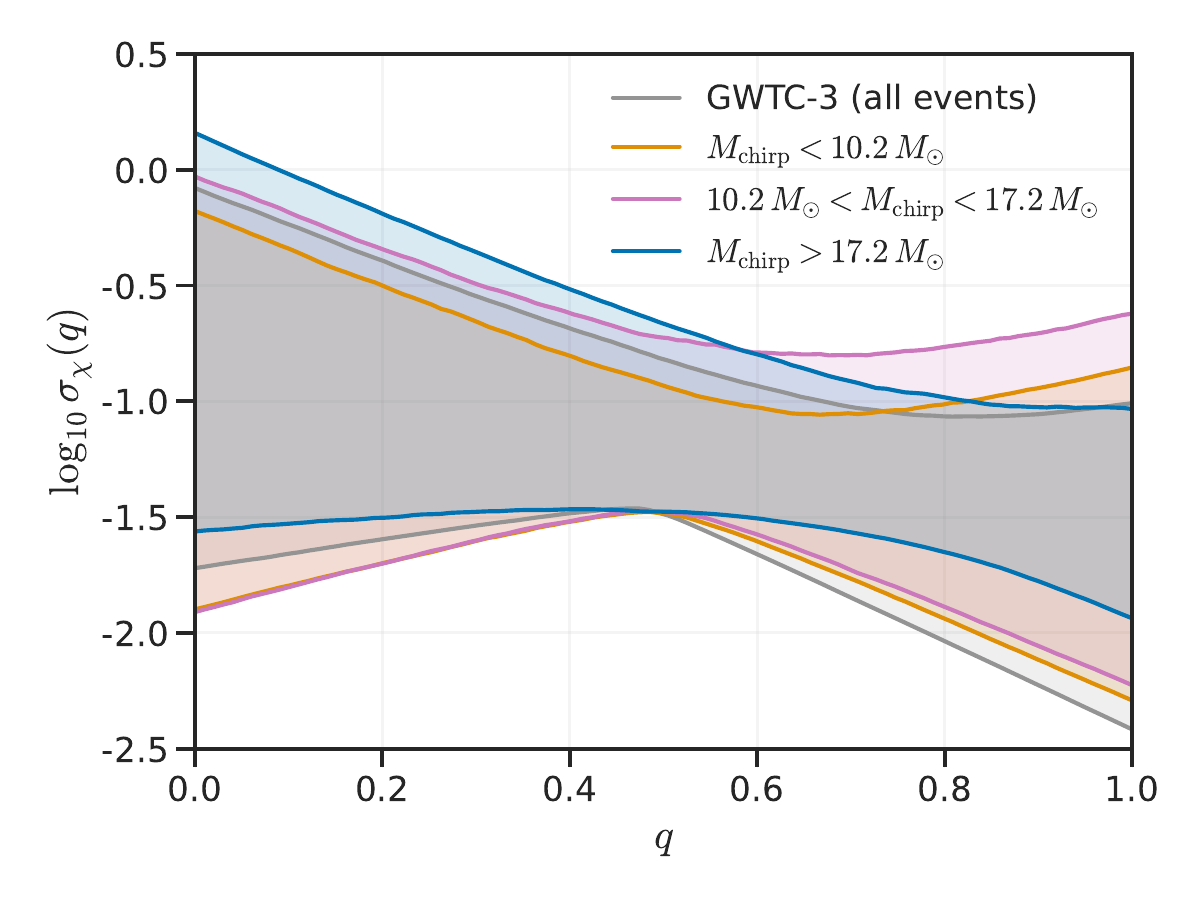}
    \caption{
        Inferred $q$-dependence of the mean effective spin, $\mu_{\chi}(q)$, (top) and width, $\sigma_{\chi}(q)$, (bottom) for each $M_{\mathrm{chirp}}$ region analyzed.
        We show the 90\% confidence intervals as the shaded regions of the $q$-$\chi_{\rm eff}$ linear relations in each of the three regions analyzed, namely, region 1 (blue), region 2 (orange) and region 3 (green) as well as the 
        the full GWTC-3 population (gray). These results illustrate how the inferred $q$-$\chi_{\rm eff}$
        correlation varies across different selections in chirp mass.
    }
    \label{fig:mu_sigma_panels}
\end{figure}

Figure~\ref{fig:mu_sigma_panels} shows the posterior distributions 90\% credible intervals of the mean effective spin, $\mu_{\chi}(q)$, and width, $\sigma_{\chi}(q)$, across all population fits considered in this work. Each colored curve corresponds to one of the chirp-mass regions defined above with the shaded bands indicating the 90\% credible intervals for the evolution of $\mu_{\chi}(q)$ and $\sigma_{\chi}(q)$. 
These results allow for a direct comparison of how the $\chi_{\rm eff}$ distribution changes when the inference is restricted to different mass domains.

\begin{figure}[t]
    \centering
    \includegraphics[width=0.45\textwidth]{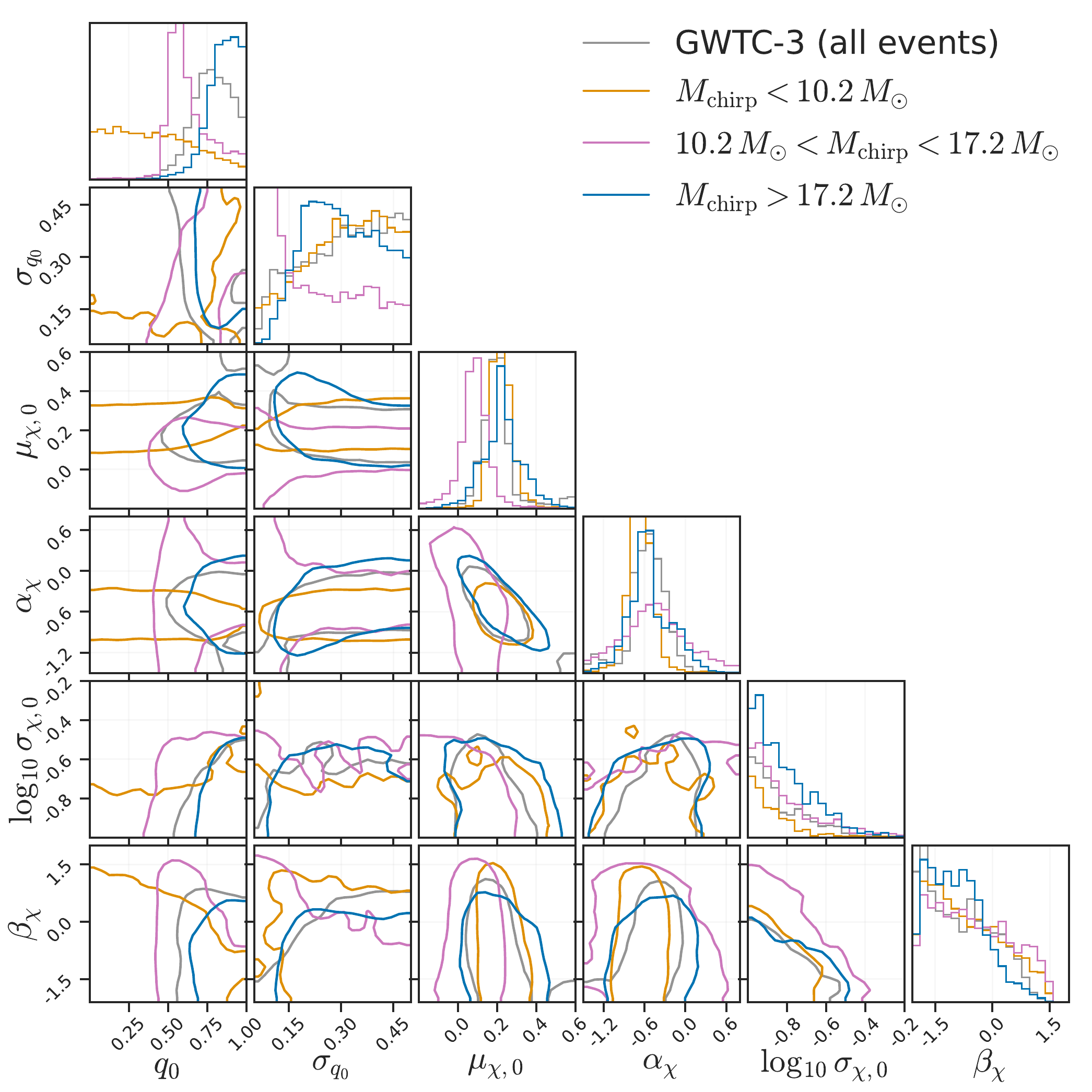}
    \caption{
        Corner plot of the model parameters that describe the linear $q$-$\chi_{\rm eff}$
        correlation model and the corresponding shape of the mass ratio distribution. Contours show the 90\% credible regions for each $M_{\mathrm{chirp}}$ region -- region 1 (blue), region 2 (orange), and region 3 (green) -- as well as the full catalog analysis (gray). The comparison highlights how the inferred intercepts and slopes vary across different $M_{\mathrm{chirp}}$ regions.
    }
    \label{fig:corner_three}
\end{figure}

Figure~\ref{fig:corner_three} presents the joint posterior distributions for the model parameters $q_0$, $\sigma_{q_0}$, $\mu_{\chi,0}$, $\alpha_{\chi}$, $\log_{10}\sigma_{\chi,0}$, and $\beta_{\chi}$ that govern the linear $q$-$\chi_{\rm eff}$ correlation model. Each contour corresponds to one of the $M_{\mathrm{chirp}}$ regions (same color scheme as in Figure~\ref{fig:mu_sigma_panels}), allowing for a direct comparison of how the inferred intercepts and slopes shift when the analysis is restricted to different subsets of the catalog. The separation or overlap of the credible regions provides a compact summary of the robustness of the inferred correlation structure and highlights which parameters are most sensitive to the choice of $M_{\mathrm{chirp}}$ interval.

Figure~\ref{fig:mu_sigma_panels} and Figure~\ref{fig:corner_three} extend the earlier discussion of how the relationship between $q$ and $\chi_{\rm eff}$ differs across the various UMAP groups. Within the adopted population-modeling framework, the top panel of Figure~\ref{fig:mu_sigma_panels} as well as the posterior distribution of $\alpha_{\chi}$ -- the slope of the $q{-}\chi_{\rm eff}$ relationship -- indicate that the lowest chirp mass region (which is identical to the Low Mass group) prefers an anti-correlation between $q$ and $\chi_{\rm eff}$: 89\% of posterior samples have $\alpha_{\chi}<0$, and the largest highest density interval (HDI) excluding $\alpha_{\chi}\geq0$ contains 83\% of the posterior mass. A comparably strong anti-correlation is also evident in the Low Mass group without imposing the population-modeling assumptions (see Figure~\ref{fig:q_chieff}). By contrast, the middle chirp mass region (which largely overlaps with the Intermediate Mass group) provides only weak constraints: 80\% of posterior samples have $\alpha_{\chi}<0$, and the HDI excluding $\alpha_{\chi}\geq0$ contains 57\% of the posterior mass. The small number of events in this region (six, as shown in Figure~\ref{fig:group_breakdown}) may limit its constraining power compared to the Low Mass group (which contains 14 events). The highest chirp mass region (which predominantly overlaps with the High Mass group) shows a preference for anti-correlation: 86\% of posterior samples have $\alpha_{\chi}<0$, and the HDI excluding $\alpha_{\chi}\geq0$ contains 75\% of the posterior probability. The indication of anti-correlation among the high-mass events is unexpected since their $q$ values are tightly clustered near unity, limiting the available domain over which an anti-correlation could be observed. We also note that Figure~\ref{fig:q_chieff} reveals no comparable trend for the High Mass group in the absence of population-modeling assumptions, suggesting that the model may induce anti-correlation for high-mass events. In summary, both with and without the population-modeling framework, we find strong evidence for an anti-correlation between $q$ and $\chi_{\rm eff}$ in the Low Mass group; for the remaining groups, the conclusions are less definitive, with weaker concordance (although no explicit disagreement) between the analyses performed with and without population-modeling assumptions.

\begin{figure}[t]
    \centering
    \includegraphics[width=\linewidth]{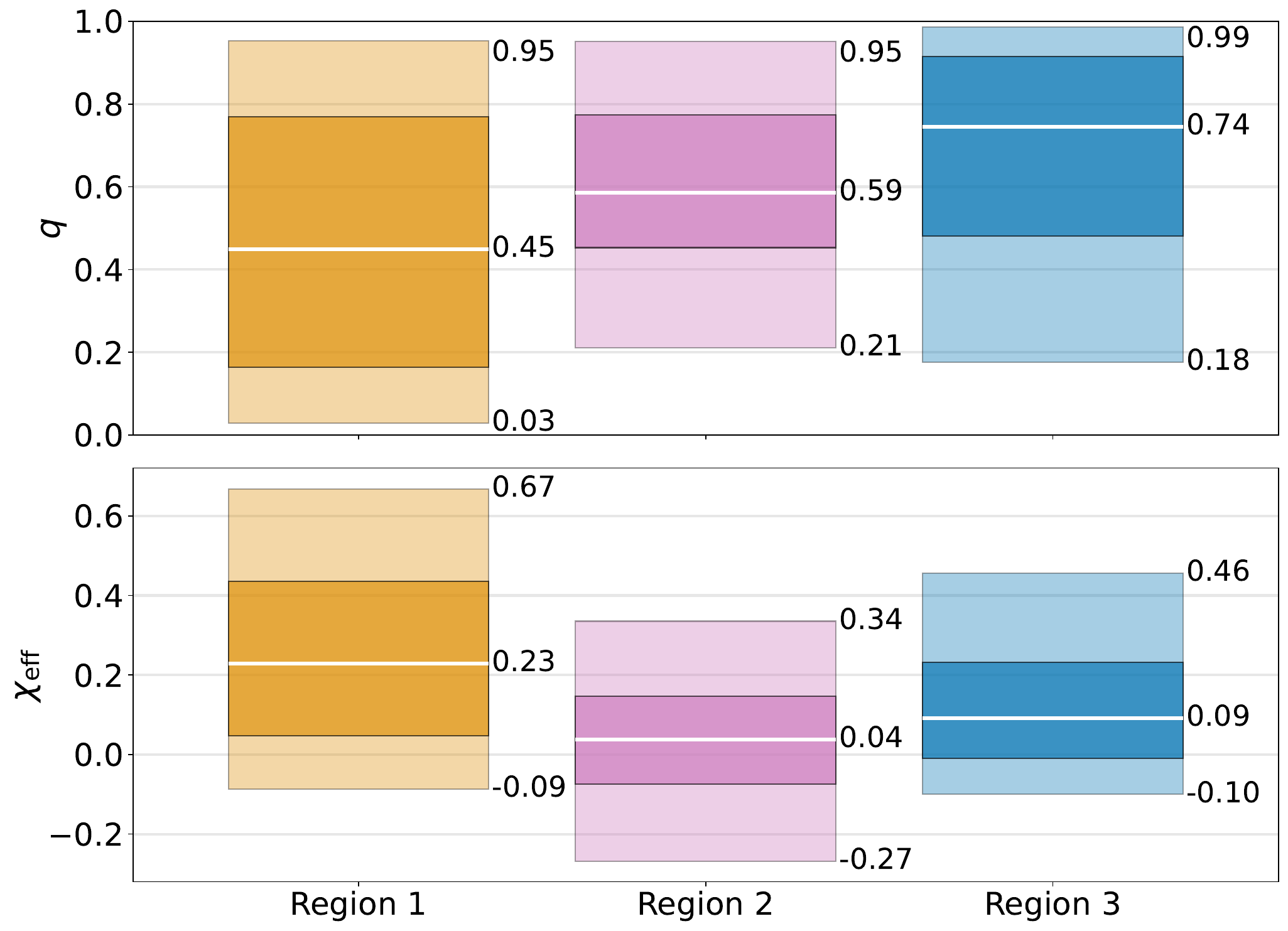}
    \caption{
        Bar plot displaying the 1$\sigma$ (less transparent) and 2$\sigma$ (more transparent) credible intervals of the marginalized probabilities $p(q)$ (top panel) and $p(\chi_{\mathrm{eff}})$ (bottom panel) for the three mass regions corresponding to our Low Mass, Intermediate Mass, and High Mass groups. Horizontal white lines indicate distribution medians. The numerical values of the medians and 2$\sigma$ bounds are annotated next to the corresponding plotted medians and 2$\sigma$ interval edges.
    }
    \label{fig:marg_post}
\end{figure}

We can extend our model parameter constraints to a characterization of the distributions of $q$ and $\chi_{\rm eff}$ in each chirp mass region through marginalization. For each of these parameters we can write the marginalization integrals as
\begin{equation}\label{eq:marg_q}
    p(q) = \int p\!\left(q \mid \Lambda_q\right)\,
           p\!\left(\Lambda_q \mid d\right)\,
           d\Lambda_q
\end{equation}
\begin{equation}\label{eq:marg_chieff}
\begin{aligned}
    p(\chi_{\mathrm{eff}})
    &= \int \int \int 
       p\!\left(\chi_{\mathrm{eff}} \mid q, \Lambda_{\chi_{\mathrm{eff}}}\right)\,
       p\!\left(q \mid \Lambda_q\right) \\
    &\quad \times
       p\!\left(\Lambda_q, \Lambda_{\chi_{\mathrm{eff}}} \mid d\right)\,
       dq \, d\Lambda_q\, d\Lambda_{\chi_{\mathrm{eff}}} \\
    &= \int \int  p(q)\,
       p\!\left(\chi_{\mathrm{eff}} \mid q, \Lambda_{\chi_{\mathrm{eff}}}\right)\,
       p\!\left(\Lambda_{\chi_{\mathrm{eff}}} \mid d\right)\,
       d\Lambda_{\chi_{\mathrm{eff}}} \, dq
\end{aligned}
\end{equation}
where $\Lambda_q = \left(q_0, \sigma_{q_0}\right)$ and $\Lambda_{\chi_{\mathrm{eff}}}=\left(\mu_{\chi,0},\, \alpha_{\chi},\, \sigma_{\chi,0},\, \beta_{\chi}\right)$. These integrals can be estimated quickly by Monte Carlo sampling the model posterior distributions ($p\!\left(\Lambda_q \mid d\right)$ for Equation~\ref{eq:marg_q} and $p\! \left(\Lambda_{\chi_{\mathrm{eff}}} \mid d\right)$ for Equation~\ref{eq:marg_chieff}) and then evaluating the remainder of the integrand from these samples. The median, 1$\sigma$, and 2$\sigma$ credible intervals (computed from the Monte Carlo samples) for each of the three chirp mass regions are shown in Figure~\ref{fig:marg_post}. We will discuss these marginalized distributions for each region in more detail in Section~\ref{section:discussion}.

\section{Discussion}\label{section:discussion}

In Section~\ref{section:intro}, we noted that UMAP's primary benefit for population analysis is avoiding the biases inherent in model-dependent methods. Using UMAP, however, presents a concomitant drawback: The algorithm's model-independence means that the results do not necessarily offer obvious astrophysical meaning. In the discussion that follows, we delineate how our UMAP results might relate to theoretical expectations of the BBH merger population.

Among the most prominent theoretical mass constraints is the prediction that instability due to pair production \citep{barkat_1967, rakavy_1967} leads to a gap in the black hole mass spectrum \citep{heger+woosley_2002}. Currently, there is little theoretical consensus on the value of the threshold mass at which this gap begins. Initial theoretical estimates ranged from ${\sim}40{-}60~ M_{\odot}$ \citep{woosley_2007, belczynski_2016_pair_ins, yoshida_2016, woosley_2019, spera_2017, leung_2019, farmer_2019}. After the discovery of higher mass mergers above these thresholds, theoretical justifications were put forth that could explain a threshold of ${\sim}70~ M_{\odot}$ \citep{woosley_2021} or even ${\sim}90~ M_{\odot}$ \citep{farmer_2020}. Others suggested that there may be a dip in the mass spectrum commencing at the threshold mass rather than a gap. In such a case the putative gap would become populated via super-Eddington accretion in isolated binaries \citep{van_son_2020, safarzadeh_2020_massgap}, dark matter annihilation in stellar progenitors \citep{ziegler_2021}, mergers in active galactic nuclei \citep{mcfacts_3, tagawa_2021}, and/or hierarchical mergers \citep{tagawa_2021, gerosa_2021, fragione_2022}. More recent analysis from GWTC-3 and GWTC-4 place the threshold at ${\sim}40{-}50~ M_{\odot}$ \citep{antonini_mass_gap, tong_mass_gap, afroz_2025} while others have argued that the threshold in GWTC-4 is closer to ${\sim}50{-}70~ M_{\odot}$ \citep{ray_rebut_mass_gap} with growing concurrence that the gap is in fact populated by hierarchical mergers. Using our fiducial UMAP output space, one might be inclined to associate the pair-instability mass threshold with the mass that delineates between the Intermediate Mass and High Mass groups. Such reasoning would lead to a threshold mass of ${\sim}30 M_{\odot}$ from $m_1$ (see Table~\ref{tab:mass_by_group}). Note, however, that a mass gap this low is found to be inconsistent with burning rates motivated by nuclear physics \citep{golomb_2024}. Perhaps a more realistic possibility is to consider $m_2$ as done in \cite{tong_mass_gap}, due to the contamination from hierarchical mergers in $m_1$. In that case, we could identify the gap with the high-end of the $m_2$ High Mass population, around ${\sim} 40~M_\odot$, close to that in \cite{tong_mass_gap}. It is also possible that the pair-instability mass threshold lies above the high-end of the high mass group ${\sim}60~M_\odot$, or that a pair-instability supernova build up delineates sub-structure within the High Mass group. Discerning such a feature would be difficult since the substructure in this group does not cluster consistently over stochastic UMAP runs, but this should be revisited with a larger number of high mass events in GWTC-4. 

Another point to stress is how the known features in the mass distribution of BBH mergers are mapped onto the UMAP space. The ${\sim}10~M_\odot$ feature in the BBH mass distribution is clearly identified with the Low Mass group, which is well separated from all other groups. The group is not as well separated in the simulations, possibly because it is drawn from the same power law distribution as the higher mass objects in that case. The ${\sim}35~M_\odot$ feature, on the other hand, comprises the High Mass group, along with higher mass objects with the exception of GW$190521\_030229$. The connection with the Intermediate Mass group and the inclusion of ${\sim}70~M_{\odot}$ objects may suggest that the ${\sim}35~M_\odot$ feature is more of a characteristic scale of a broader subpopulation rather than its own subpopulation. 

Our UMAP result did consistently find events (listed in Figure~\ref{fig:group_breakdown}) that contribute to a ${\sim}70~M_{\odot}$ peak to lie in the upper portion of the High Mass group with the exception of event GW$190521\_030229$, which we determined to be an outlier event that mostly lies on its own in the Extreme High Mass group (see Section~\ref{sub:umap_groups}). If, as suggested by \cite{ignacio_2025_bump, ignacio_2025_gp}, these events reflect a buildup of hierarchical mergers above the mass threshold (though a pair-instability supernova origin cannot be ruled out), then the separation between the High Mass and Extreme High Mass populations may illustrate UMAP’s ability to resolve distinct subpopulations within hierarchical mergers above this threshold. Previous GWTC‑2 analysis did not identify GW$190521\_030229$ as an outlier \citep{gwtc2_pop, essick_2022} although this analysis assumed a population model, which we do not assume in our UMAP-based outlier identification. Some prior analyses indicated that GW$190521\_030229$ is favored as an outlier event when specific mass population models is assumed \citep{gwtc2_pop} or once a maximum mass threshold is invoked \citep{obrien_gw190521}. We also note that the apparent outlier status of GW$190521\_030229$ may only arise once the additional GWTC‑3 events absent from GWTC‑2 are included. Future GWTC data releases may render substructures within the High Mass group more prominent, potentially allowing for more concrete placement of the mass threshold. 

Although the groups are mostly distinguished by their characteristic masses, the distribution of spins in each group can also be informative, especially when connecting the groups with potential underlying formation channels. The left lobes of the $\chi_{\mathrm{eff}}$ violin plots in Figure~\ref{fig:group_hists} show that the Low and Intermediate Mass groups peak at positive $\chi_{\mathrm{eff}}$, the Bridge and Extreme High Mass groups peak at negative $\chi_{\mathrm{eff}}$, and the High Mass group is symmetric about $\chi_{\mathrm{eff}} \approx 0$. In the population-modeling framework, these results change when reweighted by the population model as can be seen in the right lobes of the $\chi_{\mathrm{eff}}$ violin plots in Figure~\ref{fig:group_hists}. For the Low Mass and Intermediate Mass groups, the reweighted samples form a narrower distributions that peak at lower (but still positive) values of $\chi_{\mathrm{eff}}$. The High Mass group also shows narrowing. This is expected as for our choice of priors, which is typical for LVK population analyses, tends to disfavor extreme $\chi_{\mathrm{eff}}$ values while favoring values close to 0.

In Figure~\ref{fig:marg_post} we show the $p(\chi_{\mathrm{eff}})$ posteriors for the three subpopulations that closely resemble the UMAP groups. We find that the spin trend for the Low Mass group is in line with the expectation that low-mass, isolated binaries should have progenitor spins that are mostly aligned \citep{belczynski_2016_spins} leading to more positive $\chi_{\mathrm{eff}}$ \citep{kalogera_2000, farr_2017, larsen_2025}. Similarly, a high-mass, dynamical formation channel set of binaries \citep{portegies-zwart_2000} conform to an expectation of a more symmetric $\chi_{\mathrm{eff}}$ distribution, which would stem from the progenitor spins having no preferred spin orientation \citep{rodriguez_2016, talbot_2017, farr_2018}. The more negative $\chi_{\mathrm{eff}}$ groups -- the Bridge and Extreme High Mass groups -- might also result from the dynamical evolution formation channel since the lack of preferred spin orientation will result in some anti-aligned mergers, especially for mergers with $m_1 \gtrsim 20~M_{\odot}$ \citep{ye_2026}. Alternatively, the UMAP group designation may hint at a population of mergers in activate galactic nuclei \citep{mckernan_2020} or affected by spin axis tossing \citep{tauris_2022}. Another possible explanation for the Bridge samples, which tend towards primary masses of $\sim 20{-}30~M_{\odot}$, is a population of isolated binaries in which the natal kicks imparted to the binaries are preferentially either perpendicular or parallel to the spin axis; such configurations have been shown to produce anti-aligned spins in isolated binaries \citep{baibhav_antialign_2024}. Ultimately, we find that the $\chi_{\mathrm{eff}}$ distributions of the UMAP groups are consistent with different formation channels even though the primary discriminating parameter between groups is mass.

We note that left lobes in Figure~\ref{fig:group_hists} demonstrate that without population-modeling assumptions the low-mass groups have mass ratio distributions that peak at lower values ($q\sim0.4{-}0.7$ for the central 68\% of the population) when compared with the higher-mass groups ($q\sim0.6{-}0.8$). The right lobes in Figure~\ref{fig:group_hists} show that this trend of more equal mass binaries in the higher mass group persists even in the population-modeling framework. The marginalized $p(q)$ 1- and 2-$\sigma$ bounds in Figure~\ref{fig:marg_post} also show this trend of $q$ peaking at higher values for higher chirp-mass groups. Because this behavior is not found in the simulations, we are drawn to believe that it may be inherent to the data. Indeed, the latest GWTC-4 analyses also show mass ratios peaking away from unity for low-mass mergers \citep{gwtc4_pop}, claiming that this trend may conform to predictions for stable mass-transfer mergers \citep{neijssel_2019, van_son_2022}.


Although this work is the first to apply UMAP to GW subpopulation analysis, several prior studies have examined GW subpopulations using other methods, providing useful points of comparison. \cite{sridhar_2025} use a binned gaussian process to examine GWTC-4 and find three subpopulations characterized by $m_1 \lesssim 31.6 M_{\odot}$, $m_1{\sim}35~M_\odot$, and $m_1 \gtrsim 40~M_\odot$. By comparing models that explicitly assume a fixed number of subpopulations, \cite{banagiri_2025} similarly find that three subpopulations provide the most favorable characterization of GWTC‑4 with primary masses regimes of $m_1 \lesssim 28 M_{\odot}$, $28 M_{\odot} \lesssim m_1 \lesssim 40 M_{\odot}$, and $m_1>40~M_\odot$. In both these studies, the lowest mass group appears to be a combination of our UMAP-based Low Mass and Intermediate Mass groups. 
As in these studies, UMAP-based lower mass groups have $q$ distributions that peak away from unity and $\chi_{\mathrm{eff}}$ distributions with positive skew. We are intrigued by the fact that \cite{sridhar_2025} and \cite{banagiri_2025} do not distinguish the Low and Intermediate Mass groups from one another in spite of the fact that their GWTC-4 analysis includes more events than our GWTC-3 analysis, indicating the possibility that UMAP uncovers structure in GW population that cannot be found by other methods. The break at $m_1{\sim}40~M_\odot$ between two higher mass subpopulations in these studies does not appear in our UMAP analysis of GWTC-3, but we entertain the possibility that such a break would be discernible if our analysis utilized GWTC-4. Given the presence of substructure within the High Mass group -- albeit stochastically unstable -- the prospect of resolving more pronounced high‑mass features in future analyses remains promising.

We also emphasize that there are no signs of a high-mass, aligned-spins, unity-mass-ratio group, which would be strong indicators of a group produced through chemically homogeneous evolution \citep{demink_2016}. To conclude, we have shown that UMAP can be a powerful probe when attempting to connect known, theoretical expectations and/or subpopulations with GW event datasets. We expect UMAP's utility to become more apparent in future GWTC releases, which will contain more events with more parameters (beyond the fiducial four used in this work) that are well-constrained.

\section{Conclusion}\label{section:conclusion}
We use the Uniform Manifold Approximation and Projection (UMAP) algorithm as a novel, data-driven approach for exploring structure and trends within the binary black hole population of the third Gravitational-Wave Transient Catalog. Acting UMAP on individual event posterior parameter samples in a four-dimensional space (primary mass, $m_1$; secondary mass, $m_2$; effective spin, $\chi_{\mathrm{eff}}$; and redshift, $z$) results in a two-dimensional space that clearly delineates distinct subpopulations. We discern five groups mostly segregated by their component masses: Low Mass, Intermediate Mass, Bridge, High Mass, and the Extreme High Mass groups. The Extreme High Mass group is exclusively made up of samples from event GW$190521\_030229$, allowing us to conclude that this event is an outlier in the context of UMAP-based analysis. The groups show little overlap in their mass distributions at the $1\sigma$ level, confirming that mass is the primary driver of the discovered structure. We demonstrate that merely including parameters $M_{\mathrm{chirp}}$ and $\chi_{\mathrm{eff}}$ as inputs is sufficient to reproduce most of the fiducial grouping structure while the inclusion of other parameters makes group distinctions clearer and reveals finer sub-structures. Within the larger groups, the evolution of mass and spin parameters mostly follow directional axes that are approximately orthogonal to one another. We also leverage our UMAP groups, two sets of GW simulations, and a subgroups-based population model analysis to interrogate the potential anti-correlation between mass ratio, $q$, and $\chi_{\mathrm{eff}}$ in different UMAP groups. We conclude that the Low Mass group shows significant signs of anti-correlation while other groups show less significant or no signs of anti-correlation. This $q-\chi_{\mathrm{eff}}$ analysis demonstrates the potential for UMAP to reveal enlightening physical insights while not relying upon restrictive population models. Finally, we examine how our UMAP groups might relate to theoretical astrophysical expectations for binary black holes. We identify the Low Mass group with the known ${\sim}10~M_\odot$ peak whereas the ${\sim}35~M_\odot$ events do not form a distinct group but instead appear as a loose subset of the High Mass group. We also discuss the lack of a clear pair-instability mass gap in the two-dimensional UMAP space along with binary black hole formation channels that might relate to the various UMAP groups.

Since this work represents, to the best of our knowledge, the first instance of UMAP being run on GW population data, there are many avenues for further exploring UMAP's utility in a populations context. One path for further study would be to run UMAP on mergers that involve neutron star and lower mass-gap components (e.g event GW190814, which is not included in this study) -- both examining this small population on its own or in combination with binary black hole events more broadly. 
Future studies might also further investigate the tuning of UMAP hyperparameters. One might consider concocting some metric by which to judge the quality of a reduced UMAP embedding and optimize over the UMAP hyperparameter space. Such a metric might also minimize variance over stochastic UMAP runs, lending more credence to the stability and reproducibility of the UMAP results. One might also consider exploring other GW parameters as inputs in the higher-dimensional space given to UMAP. 
Finally, applying UMAP to future catalogs with more events and improved detector sensitivity may uncover additional substructure and test the robustness of the populations identified here, particularly as selection effects evolve with inception of the third-generation of gravitational wave detectors.

\section*{Acknowledgements}
The authors thank Katie Breivik, Reed Essick, and Barnabas Poczos for useful discussion. AP is supported by NSF Grant No. 2308193 and National Aeronautics and Space Administration under Grant No. 22-LPS22-0025.

This research has made use of data or software obtained from the Gravitational Wave Open Science Center (gwosc.org), a service of the LIGO Scientific Collaboration, the Virgo Collaboration, and KAGRA. This material is based upon work supported by NSF's LIGO Laboratory which is a major facility fully funded by the National Science Foundation, as well as the Science and Technology Facilities Council (STFC) of the United Kingdom, the Max-Planck-Society (MPS), and the State of Niedersachsen/Germany for support of the construction of Advanced LIGO and construction and operation of the GEO600 detector. Additional support for Advanced LIGO was provided by the Australian Research Council. Virgo is funded, through the European Gravitational Observatory (EGO), by the French Centre National de Recherche Scientifique (CNRS), the Italian Istituto Nazionale di Fisica Nucleare (INFN) and the Dutch Nikhef, with contributions by institutions from Belgium, Germany, Greece, Hungary, Ireland, Japan, Monaco, Poland, Portugal, Spain. KAGRA is supported by Ministry of Education, Culture, Sports, Science and Technology (MEXT), Japan Society for the Promotion of Science (JSPS) in Japan; National Research Foundation (NRF) and Ministry of Science and ICT (MSIT) in Korea; Academia Sinica (AS) and National Science and Technology Council (NSTC) in Taiwan.

This research used resources of the National Energy Research Scientific Computing Center (NERSC), a Department of Energy User Facility (project m4327-2025), and the Vera Cluster at the Pittsburgh Supercomputing Center (PSC).

\begin{widetext}
\appendix

\section{Hierarchical Population Inference Framework}
\label{appendix:methods}
This appendix summarizes the hierarchical Bayesian framework used to infer the model parameters $\Lambda$ governing the binary black hole (BBH) population parameter distributions.

\subsection{Default LVK Population Model}
As described in Section \ref{sub:data-gwtc3}, each detected BBH event consists of posterior samples for the source-frame primary mass $m_1$, mass ratio $q = m_2/m_1 \leq 1$, redshift $z$, and effective inspiral spin $\chi_{\mathrm{eff}}$. We denote these collectively as $\theta = (m_1, q, z, \chi_{\mathrm{eff}})$. 
 The population distribution is modeled as
\begin{equation}
    p(\theta \mid \Lambda)
    \propto
    p(m_1 \mid \Lambda)\,
    p(q \mid m_1, \Lambda)\,
    p(z \mid \Lambda)\,
    p(\chi_{\mathrm{eff}} \mid \Lambda).
\end{equation}
In what follows, we describe in more detail each of the terms in this posterior.
\paragraph{Primary mass.} The primary mass, $m_1$, is modeled using a \textsc{Powerlaw+Peak} model
\begin{equation}
    p(m_1 \mid \Lambda) = (1 - f_{\rm peak})\,
       \mathcal{P}(m_1 \mid \alpha, m_{\min}, m_{\max}, \delta m)
     + f_{\rm peak}\,
       \mathcal{N}(m_1 \mid \mu_{\rm peak}, \sigma_{\rm peak}),
\end{equation}
where $\mathcal{P}$ is a truncated power law in the range [$m_{\min}$, $m_{\max}$] with slope $\alpha < 0$. It is smoothly tapered near $m_{\min}$ using the parameter $\delta m$. The smoothed truncated power law is defined as
\begin{equation}
    \mathcal{P}(m_1 \mid \alpha, m_{\min}, m_{\max}, \delta m)
    = C m_1^{\alpha} \, S(m_1 \mid m_{\min}, \delta m)
\end{equation}
where $C$ is its normalization constant (to ensure the mixture between the power law and the peak gives a normalized primary mass distribution) and where the smoothing function $S(m_1 \mid m_{\min}, \delta m)$ is
\begin{equation}
S(m_1 \mid m_{\min}, \delta m) =
\begin{cases}
0 & \text{if } m_1 < m_{\min}, \\

1 - \exp\!\left(-\dfrac{m_1 - m_{\min}}{\delta m}\right) & \text{if } m_{\min} \leq m_1 \leq m_{\min} + \delta m, \\

1 & \text{if } m_1 > m_{\min} + \delta m.
\end{cases}
\end{equation}
The peak is implemented as a Gaussian distribution, $\mathcal{N}$, with mean $\mu_{\rm peak}$ and width $\sigma_{\rm peak}$. The mixture parameter $f_{\rm peak}$ gives the relative contribution between the smoothed power law and Gaussian components. 

\paragraph{Mass ratio.} The mass ratio $q$ is drawn from a truncated power law with slope $\beta$,
\begin{equation}
    p(q \mid \beta, m_1, m_{\min}, \delta m) \propto q^{\beta}S(m_1 \mid m_{\min}, \delta m),
\end{equation}
which is valid in the range
\begin{equation}\label{eq:qrange}
    q \in \left[\frac{m_{\min} + \delta m}{m_1},\, 1\right],
\end{equation}
ensuring consistency with the smoothing applied to the primary mass distribution. For additional details we refer the reader to Ref.~\cite{gwtc3_pop} (Appendix B, Equations~B5-6).

\paragraph{Effective inspiral spin.} The effective spin is modeled as a Gaussian with fixed mean and variance,
\begin{equation}
    p(\chi_{\rm eff} \mid \Lambda) = \mathcal{N}\!\left(\chi_{\rm eff} \mid \mu_{\chi,0}, \sigma_{\chi,0}\right),
\end{equation}
restricted to $\chi_{\rm eff} \in [-1,1]$.

\paragraph{Redshift.} The redshift evolution follows a power law, $R(z) = \mathcal{R}_0(1+z)^{\kappa}$, effectively giving a normalized redshift distribution for the BBHs, 
\begin{equation}
    p(z \mid \Lambda) \propto \frac{1}{1+z}\frac{dV_c}{dz}(1+z)^{\kappa},
\end{equation}
where $dV_c/dz$ is the differential comoving volume element and $\mathcal{R}_0$ is the merger rate at $z=0$.

\subsection{Extended Inference Model}
For the hierarchical inference that we present here, we adopt a more flexible framework that incorporates correlations and alternative parameterizations. The population distribution is modeled as
\begin{equation}
    p(\theta \mid \Lambda)
    \propto
    p(m_1 \mid \Lambda)\,
    p(q \mid m_1, \Lambda)\,
    p(z \mid \Lambda)\,
    p(\chi_{\mathrm{eff}} \mid q, \Lambda),
\end{equation}
where the effective spin distribution is explicitly conditioned on the mass ratio to allow for correlations. We retain the \textsc{Powerlaw+Peak} prescription for $m_1$ as well as the redshift model described above.

\paragraph{Mass ratio.} Instead of a power law, we use a truncated Gaussian distribution,
\begin{equation}
    p(q \mid q_0, \sigma_{q_0}, m_1, m_{\min}, \delta m) \propto \mathcal{N}(q \mid q_0,\sigma_{q_0})S(m_1 \mid m_{\min}, \delta m),
\end{equation}
which is valid in the same range as Equation~\ref{eq:qrange}. This choice provides a more flexible description for the mass ratio distribution of events in certain regions of chirp mass space.

\paragraph{Effective inspiral spin.} The effective inspiral spin distribution is conditioned on $q$, with both mean and variance allowed to evolve:
\begin{equation}
    p(\chi_{\rm eff} \mid q, \Lambda)
    = \mathcal{N}\!\left(\chi_{\rm eff} \mid \mu_{\chi}(\mu_{\chi,0},\alpha_{\chi}, q), \sigma_{\chi}(\sigma_{\chi,0},\beta_{\chi},q)\right),
\end{equation}
where
\begin{equation}
    \mu_{\chi}(\mu_{\chi,0},\alpha_{\chi}, q) = \mu_{\chi,0} + \alpha_0 (q - 0.5), \qquad
    \log_{10}\sigma_{\chi}(\sigma_{\chi,0},\beta_{\chi},q) = {\sigma_{\chi,0}} + \beta_{\chi}(q-0.5).
\end{equation}
with $\chi_{\rm eff}$ restricted to $[-1, 1]$.

This extended model provides a richer description of the BBH population, enabling inference on correlations between spin and mass ratio and offering a more realistic and flexible fit to the observed catalog when considering subpopulations.

\subsection{Rate-Marginalized Hierarchical Likelihood}

We begin with the standard hierarchical likelihood for $N_{\mathrm{det}}$ detected events, 
which includes an explicit dependence on the local merger rate $\mathcal{R}_0$:
\begin{equation}
\label{eq:likelihood_unmarg}
    \mathcal{L}\!\left(\{d_i\} \mid \Lambda, \mathcal{R}_0\right)
    \propto
    \left[\prod_{i=1}^{N_{\mathrm{det}}}
        \int 
        p\!\left(d_i \mid \theta\right)\,
        p\!\left(\theta \mid \Lambda\right)\,
        d\theta
    \right]
    \exp\!\left[-\mathcal{R}_0 \,\beta(\Lambda)\right]\,
    \mathcal{R}_0^{N_{\mathrm{det}}},
\end{equation}
where $\theta = (m_1, q, z, \chi_{\mathrm{eff}})$ denotes the source parameters, 
$p(d_i \mid \theta)$ are the single event likelihoods, and 
$\beta(\Lambda)$ is the detection efficiency under  population model $\Lambda$.

We adopt a log-uniform prior on $\mathcal{R}_0$, which allows us to marginalize the Poisson rate term explicitly:
\begin{equation}
\label{eq:ratemarg}
\mathcal{L}\!\left(\{d_i\} \mid \Lambda\right)
= \int d\mathcal{R}_0 \,
\frac{1}{\mathcal{R}_0} \,
\mathcal{L}\!\left(\{d_i\} \mid \Lambda, \mathcal{R}_0\right).
\end{equation}
This integral can be evaluated analytically, yielding a likelihood that depends only on the 
shape of the population model and the selection function~\cite{fishbach_2018}.

Explicitly, for $N_{\mathrm{det}}$ detected events with data $\{d_i\}$, the rate-marginalized population likelihood is
\begin{equation}
\label{eq:ratemarg_final}
    \mathcal{L}\!\left(\{d_i\} \mid \Lambda\right)
    \propto
    \prod_{i=1}^{N_{\mathrm{det}}}
    \frac{
        \displaystyle
        \int 
        p\!\left(d_i \mid m_1, q, z, \chi_{\mathrm{eff}}\right)\,
        p\!\left(m_1, q, z, \chi_{\mathrm{eff}} \mid \Lambda\right)\,
        dm_1\, dq\, dz\, d\chi_{\mathrm{eff}}
    }{
        \displaystyle
        \int 
        p_{\rm det}\!\left(m_1, q, z, \chi_{\mathrm{eff}}\right)\,
        p\!\left(m_1, q, z, \chi_{\mathrm{eff}} \mid \Lambda\right)\,
        dm_1\, dq\, dz\, d\chi_{\mathrm{eff}}
    }.
\end{equation}
Here $p_{\rm det}(\theta)$ is the probability of LVK detecting an event with parameters $\theta$, 
evaluated using the LVK injection campaign for BBH signals. The numerator represents the 
population-weighted evidence for event $i$, while the denominator is the selection integral, 
i.e.\ the detection efficiency $\beta(\Lambda)$ expressed explicitly in terms of $m_1, q, z,$ and $\chi_{\mathrm{eff}}$.

The detection efficiency for a given population model is
\begin{equation}
    \beta(\Lambda)
    =
    \int p_{\rm det}(\theta)\,
         p(\theta \mid \Lambda)\,
         d\theta.
\end{equation}
We estimate $\beta(\Lambda)$ via importance sampling over the injection set. The uncertainty in this estimate is modeled as a Gaussian with variance $\sigma^2 = \mu^2 / N_{\mathrm{eff}}$, where $\mu$ is the importance-sampling estimate and $N_{\mathrm{eff}}$ is the effective number of contributing. We require $N_{\mathrm{eff}} > 4 N_{\mathrm{det}}$ to avoid bias in the selection correction. For explicit details we refer the reader to Refs.~\citep{mandel_2019,farr_selection}

\subsection{Likelihood Evaluation via Importance Sampling}

The integrals in Equation~\eqref{eq:ratemarg} are evaluated using posterior samples from single-event parameter estimation. If $\{\theta_{ij}\}_{j=1}^{N_i}$ are the posterior samples for event $i$, drawn under a default prior $\pi(\theta)$, then
\begin{equation}
    \int p(d_i \mid \theta)\, p(\theta \mid \Lambda)\, d\theta
    \approx
    \frac{1}{N_i}
    \sum_{j=1}^{N_i}
    \frac{p(\theta_{ij} \mid \Lambda)}{\pi(\theta_{ij})}.
\end{equation}
The default prior used in the parameter-estimation runs is
\begin{equation}
    \pi(m_1, q, z, \chi_{\mathrm{eff}})
    \propto d_L^2 (1+z)^2 \frac{dd_L}{dz}\pi(\chi_{\mathrm{eff}} \mid \rm{PE}),
\end{equation}
where the first three factors correspond to the standard LVK priors that are uniform in comoving volume and source‑frame time. $\pi(\chi_{\mathrm{eff}} \mid \rm{PE})$ is the prior on $\chi_{\mathrm{eff}}$ induced by the parameter estimation (PE) priors assumed in LVK analyses. Because this induced prior is not analytic, we compute the corresponding $\chi_{\mathrm{eff}}$ prior using the prescription of \cite{common_priors} as implemented in the \texttt{gw-distributions} package.\footnote{\url{https://git.ligo.org/reed.essick/gw-distributions}}

\clearpage
\section{Additional UMAP results and considerations}\label{appendix:figures}

This appendix contains three figures that provide additional detail about the events in our analysis. Figure~\ref{fig:sims_35Msol} shows how simulated events associated with the ${\sim}35~M_\odot$ peak populate the UMAP embedding compared to injections outside the peak. In Figure~\ref{fig:multigroup} we show the percentage of samples that fall in each of the groups we identify for those GWTC-3 events that belong to multiple groups. In Figure~\ref{fig:group_breakdown} we show the main properties (source-frame masses, effective spin, and mass ratio) of all of the GWTC-3 events we consider, categorized by group. A tabulated version of the information contained in Figure~\ref{fig:group_breakdown} along with the parameter sample values for each event will be available on Zenodo upon acceptance.

\begin{figure*}[ht!]
  \centering
    \includegraphics[width=0.8\linewidth]{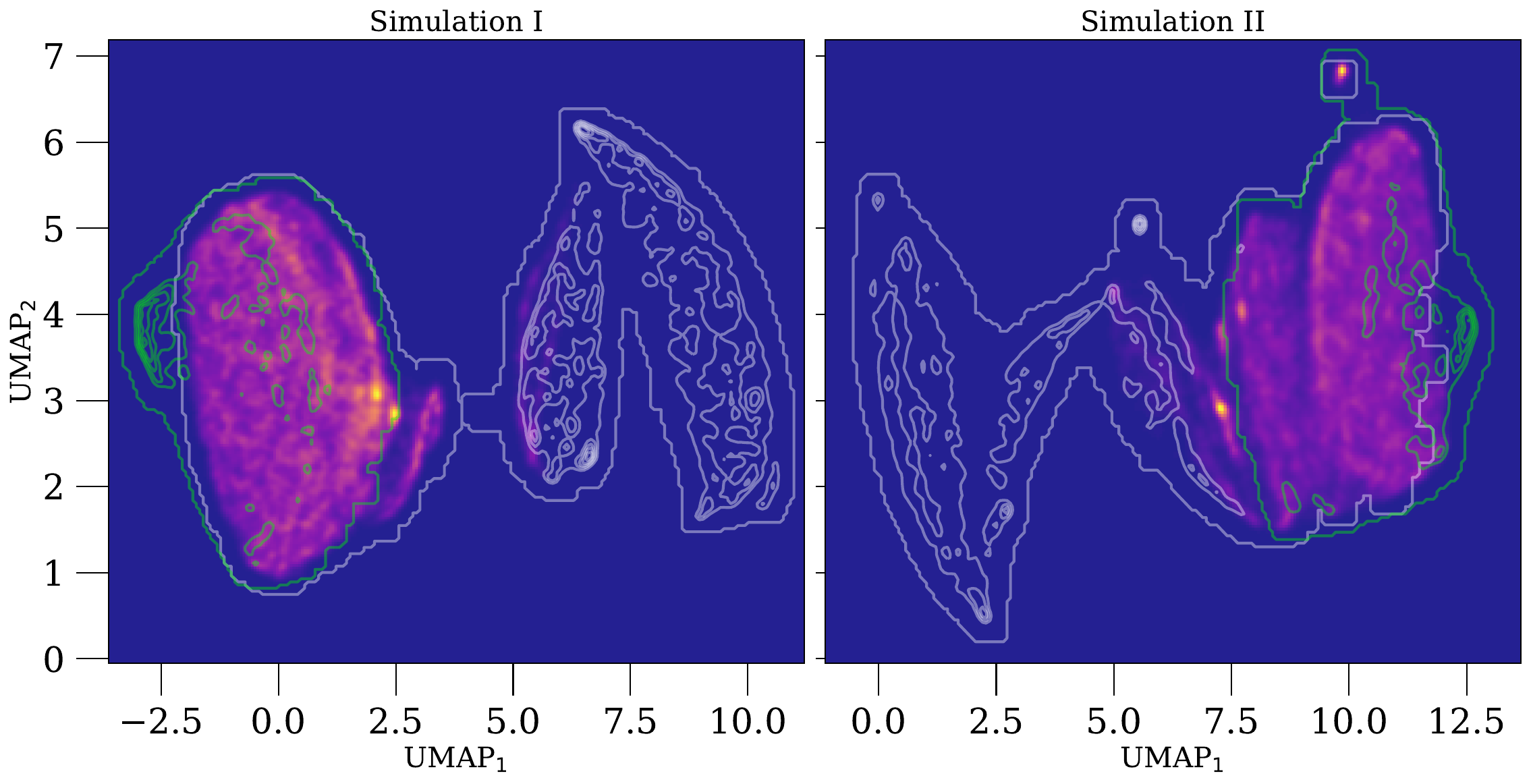}
    \caption{Smoothed density maps showing the presence of samples from simulated events in the ${\sim}35~M_\odot$ peak with primary mass $30 M_{\odot}< m_1 < 40 M_{\odot}$ compared to events outside of this regime with $m_1 < 30 M_{\odot}$ (white contours) and $m_1 > 40 M_{\odot}$ (green contours). In both cases, the contours represent five evenly spaced density levels computed from a smoothed 2D histogram of the corresponding samples. The ${\sim}35~M_\odot$ peak samples occupy a sizable portion of the High Mass group in both simulations with a subdominant presence in the High Mass group stemming from non-peak events.}\label{fig:sims_35Msol}
\end{figure*}

\begin{figure*}[ht!]
  \centering
    \includegraphics[width=0.8\linewidth]{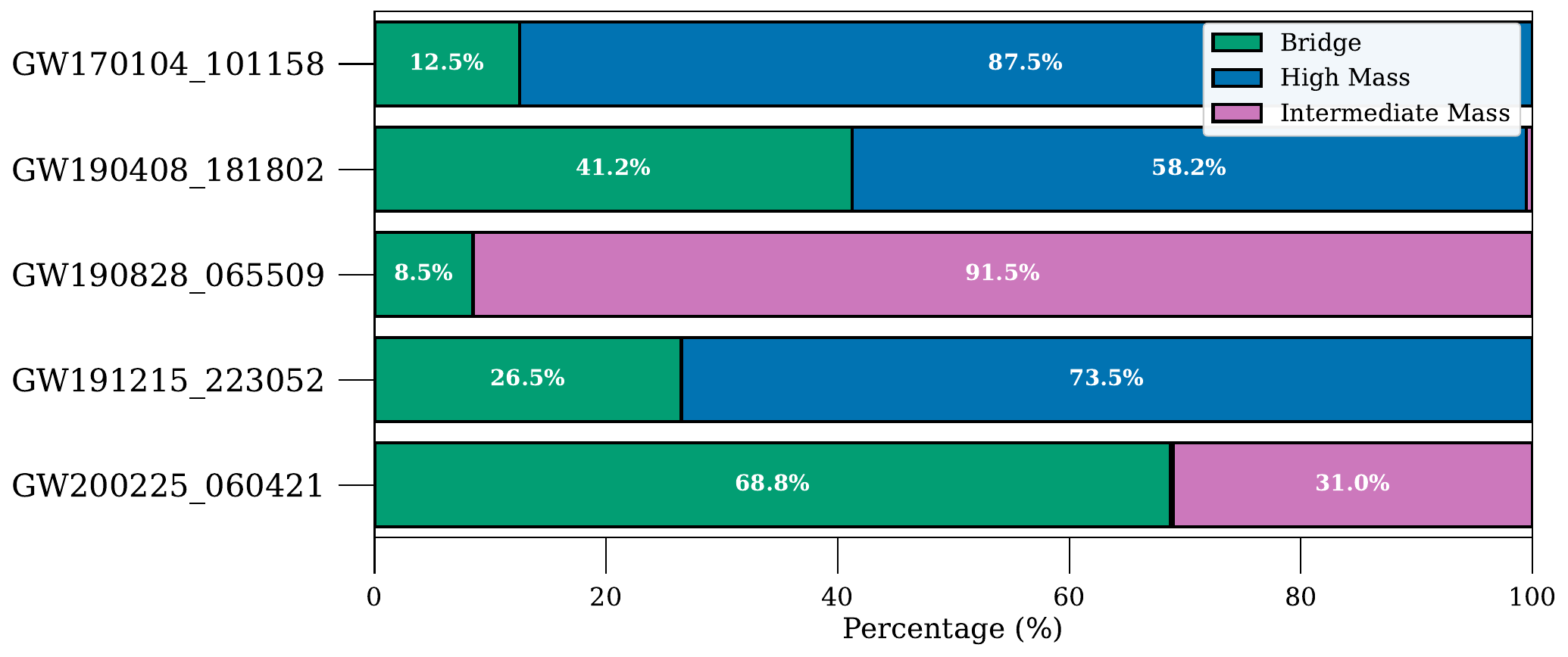}
    \caption{A breakdown of the five events for which at least 5\% of samples fall into groups other than the event’s majority group. For each event, the colored segments indicate the fraction of samples assigned to each group, with segment colors corresponding to group identity.}\label{fig:multigroup}
\end{figure*}

\begin{figure*}[ht!]
  \centering
    \includegraphics[width=\linewidth]{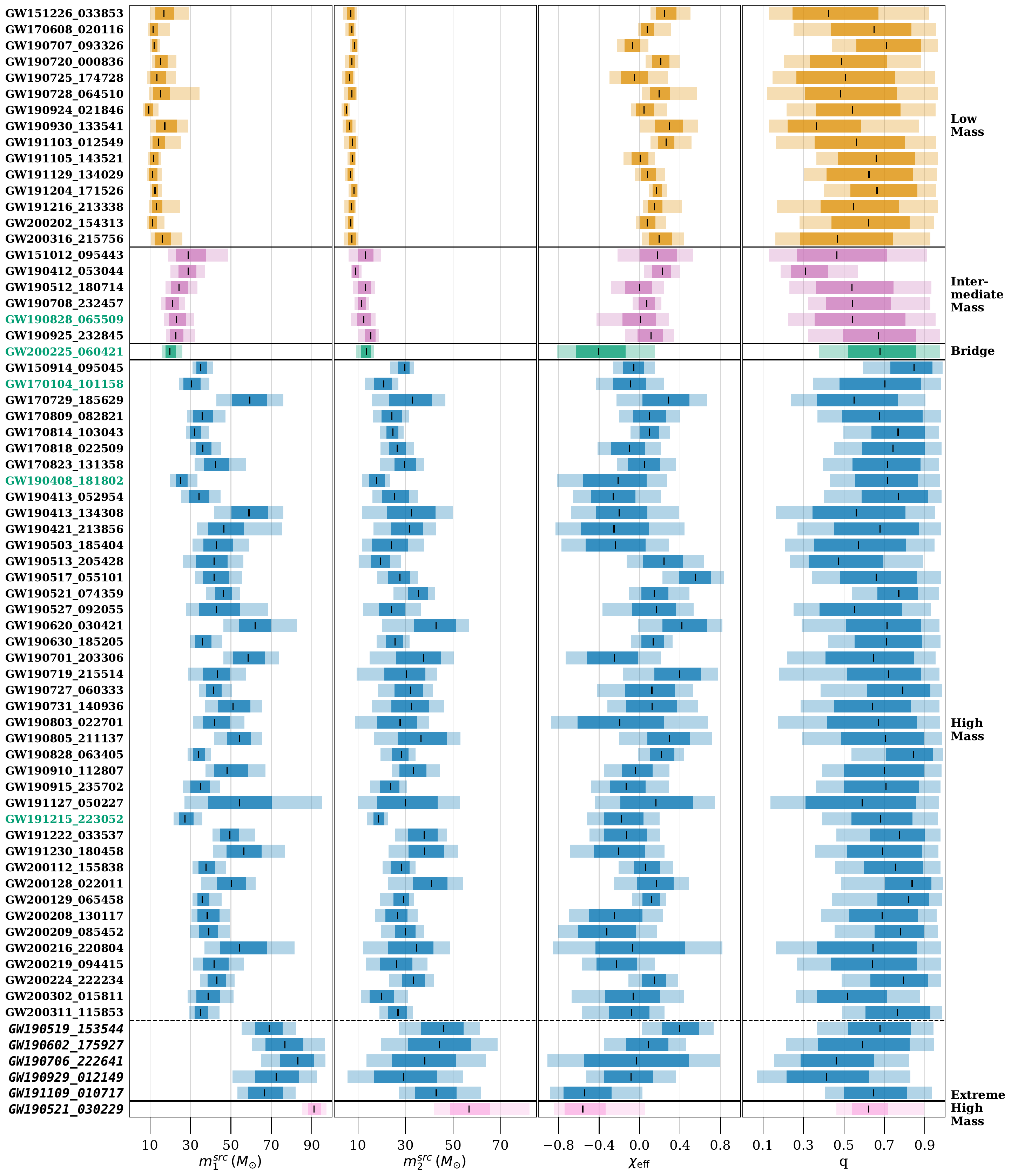}
    \caption{The 1- and 2-$\sigma$ posteriors of component masses, effective spin, and mass ratio for each GWTC-3 event. The median parameter values for each event are indicated by vertical black lines. The event rows are organized by group (from lower to higher masses) and then by date. Each event that has samples in multiple groups is associated with the group in which the plurality of its samples fall. Events with a significant number of samples in the Bridge group are given green event name labels. Events that contribute to the mass spectrum bump at ${\sim}70M_{\odot}$ (see \citep{ignacio_2025_bump}) are grouped together (below the dashed line) and have event name labels in italics.}\label{fig:group_breakdown}
\end{figure*}
\end{widetext}

\clearpage
\bibliographystyle{apsrev4-2}
\bibliography{references1}

\end{document}